\documentclass[tikz,border=5mm]{SciPost}

\hypersetup{
    colorlinks,
    linkcolor={red!50!black},
    citecolor={blue!50!black},
    urlcolor={blue!80!black}
}

\usepackage[bitstream-charter]{mathdesign}
\usepackage{romannum}
\usepackage{tikz}
\usepackage{array}
\usetikzlibrary {arrows.meta}
\usepackage{wrapfig}
\usepackage{cancel}
\usepackage{braket}
\usetikzlibrary{decorations.pathreplacing}

\DeclareSymbolFont{usualmathcal}{OMS}{cmsy}{m}{n}
\DeclareSymbolFontAlphabet{\mathcal}{usualmathcal}

\fancypagestyle{SPstyle}{
\fancyhf{}
\lhead{\colorbox{scipostblue}{\bf \color{white} ~SciPost Physics Lecture Notes }}
\rhead{{\bf \color{scipostdeepblue} ~Submission }}

\fancyfoot[C]{\textbf{\thepage}}
}

\newcommand{\pder}[2][1]{\frac{\partial #1 }{\partial #2 }} 
\newcommand*\diff{\mathop{}\!\mathrm{d}} 

\begin{document}

\pagestyle{SPstyle}

\begin{center}{\Large \textbf{\color{scipostdeepblue}{
Les Houches Lectures on Flow Networks in Biology\\
}}}\end{center}

\begin{center}\textbf{
Swarnavo Basu,
Karen Alim\textsuperscript{$\star$}
}\end{center}

\begin{center}
Technical University of Munich, Germany; TUM School of Natural Sciences, Department of Bioscience; Center for Protein Assemblies (CPA), 85748 Garching, Germany
\\[\baselineskip]
$\star$ \href{mailto:email}{\small k.alim@tum.de}
\end{center}

\section*{\color{scipostdeepblue}{Abstract}}
\textbf{\boldmath{%
Flows are essential to transport resources over large distances. As soon as diffusion becomes time-limiting, flows are needed. Flows are key for the function of multiple human organs, from the blood vasculature to the lungs, the digestive tract, the lymphatic system, and many more. While physics governs the flow dynamics, biology's response to flows governs the flow network architecture. We start with the fluid physics of Stokes flow, the prerequisite to describe the flows in biological flow networks. Then we explore how the network adaptation dynamics of biological flow networks reorganize network architecture to minimize flow dissipation or homogenize transport, storing memories of past flows along the way. 
}}

\vspace{\baselineskip}

\vspace{10pt}
\noindent\rule{\textwidth}{1pt}
\tableofcontents
\noindent\rule{\textwidth}{1pt}
\vspace{10pt}



\section{Introduction}
\label{sec:intro}
By diffusion, a molecule can cross one micrometer in one millisecond. However, it already takes 16 minutes to travel one millimeter and a painstaking 27 hours to cover one centimeter! Transport by diffusion is very slow at long distances. To transport molecules over long distances, flow networks offer fast and directed delivery. However, while any concentration gradient triggers diffusion at no cost, generating the pressure gradients that drive flows requires energy, not to mention the cost of building the tubes that channel the flows in a particular direction. Consequently, building efficient flow networks is an investment that requires careful planning, but it ultimately gives the crucial advantage of outpacing diffusion.

Flow networks are a fundamental building block in biology. They are replicated multiple times in our organs, including the vasculature, the digestive tract, the lungs, the lymphatic system, and many more. Entire organisms, like fungal networks and slime molds, are also structured as flow networks. 
These networks have to be designed to ensure functionality while simultaneously reducing costs. When it comes to designing networks, biology fundamentally differs from an engineer building a flow network. First, there is no global design, so adjustments to the network architecture can only happen locally based on locally available information. Second, we do not know the overall budget plan of a living being and how each investment is balanced by whatever function it achieves. Yet, the search for principles that constrain the architecture of biological flow networks by their cost, function, and adaptability has identified major knobs that determine the network architecture. 

Investigating flow networks in biology is mesmerizing for the artistic eye marveling at the beautiful patterns emerging from life. 
It involves complex physics that is intellectually challenging as flows are globally coupled in a flow network, and it uncovers the workings and malfunctions of vital organs, like the vasculature.
These lecture notes are aimed to guide the reader into the general principles of flow networks in biology by spotlighting concepts and ideas in specific examples rather than giving an exhaustive review of the literature. The examples selected are intended to foster physical intuition in the reader by illustrating concepts with analytical derivations and formulas.
Sadly, these notes cannot reflect the vivid discussions that took place in the lecture hall beyond the utterly picturesque setting facing Mont Blanc at Les Houches. 

We start in chapter 2 with a primer on the fluid physics of flows in tubes. We derive the Navier-Stokes equations of incompressible, Newtonian fluids; discuss flow in a cylindrical tube and the energy dissipated by the flow. Then, we couple the tubes and their flows into a network of inter-connected tubes. In chapter 3, we introduce how the concept of minimizing the energy dissipated by fluid flow at a finite metabolic cost motivates tube radius adaptation in response to local flow shear stress. We derive shear stress adaptation from force balance and map out how such adaptation shapes network architecture. In chapter 4, we connect the architecture of adaptive flow networks to disordered systems and non-neural information processing by exemplifying how flow shear stress adaptation stores memories of past flow in the network architecture. Finally, in chapter 5, we branch out to flow network functions and how functions constrain network architecture by discussing distributed production networks and transport of resources with flow, before concluding these Lecture Notes.
\section{Flows in tubes and networks}
As fluid physics governs the dynamics of flows in flow networks, we start with a refresher on fluid dynamics tailored to tubes and tubular networks. The basis of fluid physics is the Navier-Stokes equation, which we derive and solve to obtain the flow in a cylindrical tube, known as Poiseuille flow. We move on to determine a flow property of particular interest, the flow shear stress, which is instrumental in quantifying the energy dissipated during flow. Finally, connecting tubes into a network, we derive how flows in individual tubes are coupled together via conservation of fluid volume.
\subsection{Navier-Stokes Equations}
A fluid is well described as a continuum characterized by its density field $\rho(\mathbf{x},t)$ and its velocity field $\mathbf{u}(\mathbf{x},t)$ in space $\mathbf{x}$ and time $t$. The dynamics of the fluid are governed by the continuity equation and the Navier-Stokes equation, which together provide four equations for the four unknowns, the density field and the three components of the velocity field. 

The continuity equation describes mass conservation. In a fluid element, density changes due to the in- and out-flux of fluid of density $\rho$, which can be mathematically written as
\begin{align}\label{eq:MassCons}
    \frac{\partial \rho}{\partial t} + \mathbf{\nabla} \cdot (\rho \mathbf{u}) = 0,
\end{align}
where $\mathbf{\nabla}$ denotes the gradient operator. If the fluid is incompressible, its density is independent of space and time, and Eq.~\eqref{eq:MassCons} simplifies to
\begin{align}\label{eq:MassConsInc}
    \frac{\partial \rho}{\partial t} + \mathbf{\nabla} \cdot (\rho \mathbf{u}) &= \underbrace{\frac{\partial \rho}{\partial t} + \mathbf{u}\cdot(\mathbf{\nabla} \rho)}_{\text{$=0$ (incompressibility)}} + \rho (\mathbf{\nabla}\cdot \mathbf{u}) = 0,
\end{align}
which gives the continuity condition for incompressible fluids,
\begin{align}\label{eq:Incompressibility}
\mathbf{\nabla}\cdot \mathbf{u} &=0, 
\end{align}
stating that incompressible flows are divergence-free.

The Navier-Stokes equation follows from the balance of forces acting on a fluid element: 
\begin{align}
    \underbrace{\frac{d(\rho \mathbf{u})}{dt}}_{\text{rate of change of momentum}}=\underbrace{\vphantom{\frac{d(\rho \mathbf{u})}{dt}}\mathbf{f_{\rm b}}}_{\text{external body forces}}+\underbrace{\vphantom{\frac{d(\rho \mathbf{u})}{dt}}\nabla\cdot\sigma}_{\text{internal forces due to deformation}}
\end{align}
where $\mathbf{f_{\rm b}}$ is the sum of all external forces per unit volume and $\sigma$ is the stress tensor of the fluid. For incompressible fluids, the left-hand side of the above expression simplifies as follows:
\begin{align}
    \frac{d(\rho \mathbf{u})}{dt} &= \pder[(\rho\mathbf{u})]{t}+\pder[(\rho\mathbf{u})]{x_i}\pder[x_i]{t}, \qquad \text{ (using Einstein's summation convention)}\nonumber \\ 
    &=\mathbf{u}\underbrace{\pder[\rho]{t}}_{\Romannum{1}}+\rho\pder[\mathbf{u}]{t}+\pder[x_i]{t}\left( \rho\pder[\mathbf{u}]{x_i} + \underbrace{\mathbf{u}\pder[\rho]{x_i}}_{\Romannum{2}}\right), \nonumber \\ 
    &=\underbrace{\mathbf{u}\left(\underbrace{\pder[\rho]{t}}_{\Romannum{1}}+\underbrace{\pder[x_i]{t}\pder[\rho]{x_i}}_{\Romannum{2}}\right)}_{\text{$=0$ (incompressibility)}}+\rho\pder[\mathbf{u}]{t}+\rho\pder[x_i]{t} \pder[\mathbf{u}]{x_i}, \nonumber \\
    &=\rho\pder[\mathbf{u}]{t}+\rho\pder[x_i]{t}\pder[\mathbf{u}]{x_i},  \nonumber\\ 
    &= \rho\pder[\mathbf{u}]{t}+\rho (\mathbf{u}\cdot\nabla)\mathbf{u}.
\end{align}
For Newtonian fluids, the stress tensor is defined as
\begin{align}\label{NewtonianStress}
    \sigma=-P \mathbb{I} +\underbrace{\mu\left(\nabla \mathbf{u} +(\nabla \mathbf{u})^T \right)}_{\text{shear stress, $\Tilde{\tau}$}},
\end{align}
where $P$ is the pressure field, $\mathbb{I}$ is the identity matrix and $\mu$ is the viscosity of the fluid. Using the definition of the stress tensor, the internal forces due to deformation are written as
\begin{align}
    \nabla\cdot\sigma&=-\nabla P+\nabla\mu\left(\nabla\mathbf{u}+(\nabla\mathbf{u})^T\right), \nonumber \\
    &=-\nabla P+\mu \nabla^2\mathbf{u}+\underbrace{\mu\nabla(\nabla\cdot\mathbf{u)}}_{\text{$=0$ (Eq. \eqref{eq:Incompressibility})}}, \nonumber \\
    &=-\nabla P+\mu\nabla^2\mathbf{u} .
\end{align}
Combining all the terms together again, one obtains the Navier-Stokes equation for incompressible, Newtonian fluids:
\begin{align}\label{eq:NavierStokes}
   \underbrace{\rho\pder[\mathbf{u}]{t}}_{\text{steadyness}} + \underbrace{\vphantom{\pder[\mathbf{u}]{t}}\rho(\mathbf{u\cdot\nabla) u}}_{\text{inertia}}=\underbrace{\vphantom{\pder[\mathbf{u}]{t}} \mathbf{f_{\rm b}}}_{\text{body forces}}-\underbrace{\vphantom{\pder[\mathbf{u}]{t}}\mathbf{\nabla}P}_{\text{pressure force}}+\underbrace{\vphantom{\pder[\mathbf{u}]{t}}\mu\nabla^2\mathbf{u}}_{\text{viscous force}}.
\end{align}
Non-dimensionalizing the Navier-Stokes-Eq.~\eqref{eq:NavierStokes} with non-dimensional variables $\mathbf{u^*}=\mathbf{u}/U$, $t^*=t/T$, $\mathbf{x^*}=\mathbf{x}/L$, $p^*=P/(\rho U^2)$, and assuming $\mathbf{f_{\rm b}}=0$ identifies two non-dimensional numbers that govern fluid dynamics, the Strouhal number  $\mathrm{St}=L/(UT)$ and the Reynolds number $\mathrm{Re}= (UL)/\gamma$, where $\gamma=\mu/\rho$ is the kinematic viscosity. The Navier-Stokes-Eq.~\eqref{eq:NavierStokes} in non-dimensional form is thus written as
\begin{align}\label{eq:NavierStokesNonDim}
    \mathrm{St}\pder[\mathbf{u^*}]{t^*}+(\mathbf{u^*\cdot\nabla})\mathbf{u^*} = -\mathbf{\nabla}p^* + \frac{1}{\mathrm{Re}}\nabla^2\mathbf{u^*}.
\end{align}
 For small Reynolds number, $\mathrm{Re}\ll1$, the advective force is much smaller than the viscous force, and the Navier-Stokes equation Eq.~\eqref{eq:NavierStokes} simplifies to the Stokes equation,
\begin{align}\label{eq:StokesFlow}
    \pder[\mathbf{u}]{t}=-\frac{1}{\rho}\mathbf{\nabla}P+\gamma\nabla^2\mathbf{u}.
\end{align}
Due to their small dimensions and their low flow velocities, flow networks in biology typically fall within the realm of small Reynolds numbers. Therefore, the flow velocity field $\mathbf{u}(\mathbf{x},t)$ and pressure $P$ follow from the Stokes-Eq.~\eqref{eq:StokesFlow} and the incompressibility condition, Eq.~\eqref{eq:Incompressibility}, subject to the boundary conditions. For flow in flow networks, the basis is the flow in a cylindrical tube.
\subsection{Flow in a tube - Poiseuille flow}
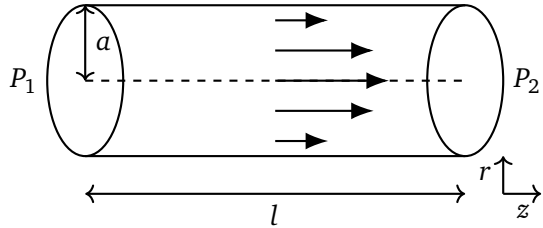
\begin{wrapfigure}{r}{0.5\textwidth}
    \begin{tikzpicture}
    \def\a{1} 
    \def\l{5} 


    \draw[thick,, dashed] (0, 0) -- (\l, 0);
    \draw[thick] (0, \a) -- (\l, \a);
    \draw[thick] (0, -\a) -- (\l, -\a);

    \draw[thick] (0, 0) ellipse[x radius=0.5, y radius=\a];
    \draw[thick] (\l, 0) ellipse[x radius=0.5, y radius=\a];

    \draw[-{Latex[length=3mm]}, thick] (2.5, 0) -- ++(1.5, 0); 
    \draw[-{Latex[length=3mm]}, thick] (2.5, 0.4) -- ++(1.3, 0); 
    \draw[-{Latex[length=3mm]}, thick] (2.5, -0.4) -- ++(1.3, 0); 
    \draw[-{Latex[length=3mm]}, thick] (2.5, 0.8) -- ++(0.7, 0); 
    \draw[-{Latex[length=3mm]}, thick] (2.5, -0.8) -- ++(0.7, 0); 

    \node[anchor=east] at (-0.5, 0) {$P_1$};
    \node[anchor=west] at (\l+0.5, 0) {$P_2$};

    \draw[thick, ->] (\l+0.5, -1.5) -- ++(0.5, 0) node[below left] {$z$};
    \draw[thick, ->] (\l+0.5, -1.5) -- ++(0, 0.5) node[below left] {$r$};

    \draw[<->, thick] (0, 0) -- (0, \a) node[midway, right] {$a$};

    \draw[<->, thick] (0, -1.5) -- (\l, -1.5) node[midway, below] {$l$};
    
\end{tikzpicture}      
    \caption{\textbf{Flow profile in a cylindrical tube.} In a cylindrical tube of radius $a$, length $l$ the pressure drop $\Delta P = P_1-P_2$ drives a parabolic flow profile known as Poiseuille flow.}
    \label{fig:Poiseuille}
\end{wrapfigure}
Consider a cylindrical tube of radius $a$ and length $l$ extending along the longitudinal axis $z$, and radial axis $r$, see Fig.~\ref{fig:Poiseuille}. At the wall, the flow velocity decays to zero, a so-called no-slip boundary condition,  i.e.~$\mathbf{u}(r=\pm a,t)=0$. Solving the Navier-Stokes-Eq.~\eqref{eq:NavierStokes} for an incompressible Newtonian fluid at steady state velocity yields
\begin{align}\label{eq:PoiseuilleFlow}
    u_z=-\frac{(P_2-P_1)a^2}{4\mu l}\left(1-\frac{r^2}{a^2}\right),
\end{align}
and zero velocity in both the angular and radial directions. 
The cross-sectionally averaged flow velocity, $\bar{u}$, then follows as:
\begin{align} \label{eq:FlowVelHagenPoiseuille}
    \bar{u} &= \frac{1}{\pi a^2}\int_0^a -\frac{(P_2-P_1)a^2}{4\mu l}\left(1-\frac{r^2}{a^2}\right) 2\pi r \diff{r}, \nonumber \\
   \implies \bar{u} &=-\frac{(P_2-P_1)a^2}{8\mu l}.
\end{align}
Defining the flow rate, $Q$, as the amount of fluid passing per unit time through a cross-sectional area, $A$, relates the flow rate $Q$ to the pressure drop $\Delta P = (P_1-P_2)$ by the following relation, known as the Hagen-Poiseuille law:
\begin{align}\label{eq:ResHagenPoiseuille}
    Q&=\bar{u}A=\frac{\Delta P \pi a^4}{8\mu l}, \nonumber \\
    \implies \Delta P &= \frac{8\mu l}{\pi a^4}Q = RQ = \frac{Q}{C},
\end{align}
where we define $R=\frac{1}{C}=\frac{8\mu l}{\pi a^4}$. The pressure drop acts like a gradient in a potential that drives the motion of fluid volume. Flow in a tube and tubular networks thereby parallels charge transport in electric wires and circuits, see Table~\ref{tab:ElectroHydro}. In fact, analogous to electrical circuits, we identify the hydraulic resistance for fluids as the proportionality constant $R$ that relates the pressure drop $\Delta P$ and the flow rate $Q$ as $\Delta P = RQ$. Consequently, hydraulic conductance is defined as the inverse of hydraulic resistance $C=1/R$. For a cylindrical tube with Poiseuille flow, $R$ is given by $R=\frac{8\mu l}{\pi a^4}$. The notion that a fluid resists motion is tied to the shearing of the fluid, which arises when adjacent layers move past one another. The resulting shear stresses dissipate the energy invested to move the fluid.
\begin{table}[h]
\centering
\begin{tabular}{|l|c|c|}
\hline
 & \textbf{Electric wire} & \textbf{Hydraulic tube} \\\hline
Quantity & $q'$, charge & $V$, volume \\\hline
Quantity flux & $I$, current & $Q$, flow rate \\\hline
Potential & $U$, electric potential & $P$, pressure \\\hline
Linear model & $I=\frac{\Delta U}{R'}$, Ohm's law & $Q=\frac{\pi a^4}{8\mu l}\Delta P$, Poiseuille flow \\\hline
Resistance & $R'$, electric resistance & $R=\frac{8\mu l}{\pi a^4}$, hydraulic resistance \\\hline
Conductance & $C'=1/R'$ & $C=\frac{1}{R}=\frac{\pi a^4}{8\mu l}$ \\\hline
\end{tabular}
\caption{Electric-Hydraulic analogy}
\label{tab:ElectroHydro}
\end{table}
\subsection{Shear stress in Poiseuille flow}
To quantify the dissipation caused by friction between layers of fluid moving past each other, we need to quantify the shear stress. For a flow of cylindrical symmetry, as in a tube, see Fig.~\ref{fig:Poiseuille}, the individual components of the shear stress tensor $\Tilde{\tau}$ of the Newtonian fluid, see Eq.~\eqref{NewtonianStress}, are in cylindrical coordinates $r$, $z$, $\theta$, defined as
\[
\begin{aligned}
&\tau_{rr} = 2\mu \frac{\partial u_r}{\partial r},
&\tau_{\theta\theta} = 2\mu \left( \frac{1}{r} \frac{\partial u_\theta}{\partial \theta} + \frac{u_r}{r} \right),\\
&\tau_{zz} = 2\mu \frac{\partial u_z}{\partial z},
&\tau_{r\theta} = \tau_{\theta r} = \mu \left( \frac{\partial u_r}{\partial \theta} + \frac{\partial u_\theta}{\partial r} - \frac{u_\theta}{r} \right),\\
&\tau_{rz} = \tau_{zr} = \mu \left( \frac{\partial u_r}{\partial z} + \frac{\partial u_z}{\partial r} \right),
&\tau_{\theta z} = \tau_{z\theta} = \mu \left( \frac{1}{r} \frac{\partial u_z}{\partial \theta} + \frac{\partial u_\theta}{\partial z} \right).
\end{aligned}
\]
Note that by definition, $\Tilde{\tau}$ is symmetric, see Eq.~\eqref{NewtonianStress}. For Poiseuille flow, the only non-zero elements in the shear stress tensor are
$$ \tau_{rz} = \tau_{zr} = \mu \left( \frac{\partial u_r}{\partial z} + \frac{\partial u_z}{\partial r} \right) = \mu  \frac{\partial u_z}{\partial r} = \frac{(P_2-P_1)r}{2l} \quad (\text{employing Eq.~\eqref{eq:PoiseuilleFlow}}).$$ 
Consistent with Hagen-Poiseuille flow, fluid is only moving along the tube along $z$, and thus fluid layers are only sheared due to their different velocities in radial directions. The larger the radial coordinate of the fluid layer, the larger the difference between the velocities of adjacent flow layers and the greater the shear stress. The shear stress is the largest at the tube wall. Expressing the shear stress in terms of the flow rate,  see Eq.~\eqref{eq:ResHagenPoiseuille}, the maximal shear stress acting on the tube wall $\tau$ is
\begin{align}\label{eq:WallStress}
    \tau=-\tau_{rz} \big|_{r=a} = \frac{4Q\mu}{\pi a^3}.
\end{align}
This wall shear stress plays an important role in biological flow networks, as cells or semiflexible polymer-gels making up the tube wall may respond to the shear stress and drive the adaptation of tube radius. The shear stress within the entire cross-section of the tube contributes to the overall dissipation of energy due to friction between fluid layers.
\subsection{Energy dissipation in Poiseuille flow}
To quantify the energy dissipated by flow of incompressible fluids, we derive the rate of energy loss in the fluid per unit volume, known as energy dissipation, by calculating the rate of change of kinetic energy in a fluid volume V as follows:
\begin{align}
    \frac{d}{dt}\int_V \diff{V} E_{\text{kin}}&=\int_V\diff{V}\left(\pder[E_{\text{kin}}]{t}+(\mathbf{u}\cdot\nabla)E_{\text{kin}}\right), \nonumber \\
    &=\int_V\diff{V}\left(\pder[E_{\text{kin}}]{t}+ \nabla\cdot(E_{\text{kin}}\mathbf{u})-\underbrace{E_{\text{kin}}\nabla\cdot\mathbf{u}}_{\text{=0 (incompressible)}}\right), \nonumber \\
    &=\int_V\diff{V}\pder[E_{\text{kin}}]{t}+\int_S E_{\text{kin}}\mathbf{u}\diff{\mathbf{S}},\nonumber
\end{align}
where we employed Gauss's theorem to transform the volume integral into a surface integral. Now the surface integral over the kinetic energy times the flow velocity evaluates to zero if we consider the volume to be infinite, or if we assume no-slip boundary conditions. Therefore, the total derivative of the kinetic energy is expressed as a partial derivative of the kinetic energy  as the integrand, 
\begin{align}\label{eq:KinEnergy}
    \frac{d}{dt}\int_V \diff{V} E_{\text{kin}}=\int_V\diff{V}\pder[E_{\text{kin}}]{t}.
\end{align}
The partial derivative of the kinetic energy can be written as
\begin{align}
    \pder[E_{\text{kin}}]{t}=\pder[]{t}\left(\frac{\rho u^2}{2}\right)=\rho \mathbf{u}\cdot\pder[\mathbf{u}]{t}.
\end{align}
Now we employ the Navier-Stokes-Eq.~\eqref{eq:NavierStokes} for incompressible, Newtonian fluids and solve for a fluid that is not subjected to any external forces, to obtain
\begin{align}
    \pder[E_{\text{kin}}]{t}=\rho u_i\pder[u_i]{t}&=u_i\left( -u_j\partial_ju_i-\partial_i P+\partial_j\tau_{ji}\right), \nonumber \\
    &=-u_j\partial_j\left(\frac{u_i u_i}{2}+P\right)+u_i\partial_j\tau_{ji}, \nonumber \\
    &= -\partial_j\left[ u_j\left(\frac{u_i u_i}{2}+P\right) \right]-\underbrace{\left(\frac{u_i u_i}{2}+P\right)\partial_j u_j}_{\text{=0 (incompressible)}} + \partial_j(\tau_{ji}u_i)- \tau_{ji}\partial_j u_i, \nonumber \\
    &=\partial_j\left[- u_j\left(\frac{u_i u_i}{2}+P\right) + \tau_{ji}u_i \right] - \tau_{ji}\partial_j u_i, \nonumber
\end{align}
where we again used Einstein summation. Plugging the result for the partial derivative of the kinetic energy back into the integral, we again use Gauss's theorem to write the total rate of change of kinetic energy Eq.~\eqref{eq:KinEnergy} as
\begin{align}
    \int_V\diff{V}\pder[E_{\text{kin}}]{t}&=\int_s\left[- u_j\left(\frac{u_i u_i}{2}+P\right) + \tau_{ji}u_i \right]\diff{\mathbf{S}}-  \int_V\diff{V}\left(\tau_{ji}\partial_j u_i \right).
\end{align}
As before, the surface integral vanishes to zero if one extends the surface to infinity where the velocities vanish, or if one assumes no-slip boundary conditions at the surface. Therefore, the total rate of change of kinetic energy, i.e. the  energy dissipation, reduces to
\begin{align}\label{eq:Dissipationgen}
      \frac{d}{dt}\int_V \diff{V} E_{\text{kin}}=\int_V\diff{V}\pder[E_{\text{kin}}]{t}= -\int_V\diff{V}\left( \tau_{ji}\partial_j u_i \right).
\end{align}
For Poiseuille flow in a cylinder of radius $a$ and length $l$, Eq.~\eqref{eq:FlowVelHagenPoiseuille},  only the shear stress $\tau_{zr}=\tau_{rz}$ is non-zero, such that the energy dissipation  Eq.~\eqref{eq:Dissipationgen} evaluates to
\begin{align}
\label{eq:Dissipation}
    \frac{d}{dt}\int_V \diff{V} E_{\text{kin}}= - \int_0^a 2 \pi lr \diff{r} \tau_{zr}\partial_ru_z = -\int_0^a 2 \pi lr \diff{r} \left[\mu\left( \frac{\Delta Pr}{2\mu l}\right)^2\right] = -Q\Delta P = -\frac{Q^2}{C}.
\end{align}
Therefore, energy dissipation increases with flow rate $Q$, but can be compensated for by increasing the tube's conductance, $C$. If possible, the energy dissipation would be minimized by increasing the tube's conductance to infinity, which would, for example, correspond to an infinitely large tube radius $a$. This balance between flow rate and tube conductance will be instrumental when investigating flow networks that minimize energy dissipation. 
\subsection{Flows in networks}
To set up the equations for the flows in networks, we build on the electric-hydraulic analogy, which extends beyond the single wire-tube analogy to entire networks. In particular, analogous to the conservation of charge, the fluid volume is conserved at every network node, known as Kirchhoff's first law. In addition, Kirchhoff's second law states that the summed pressure drop along a closed circuit is zero. To define Kirchhoff's laws, we label network nodes by $i\in N$ and refer to individual tubes $ij$ by the nodes they connect, i.e.~node $i$ and node $j$. Thus, the flow rate $Q_{ij}$ in a tube $ij$ is given by $Q_{ij}=C_{ij}(P_j - P_i)$, where $C_{ij}$ is the hydraulic conductance of the tube and $P_i$ and $P_j$ are the pressures at the nodes $i$ and $j$, respectively. In general, the tube conductance $C_{ij}$ depends on the tube geometry and its boundary conditions which determine the flow profile in the tube. For Poiseuille flow in cylindrical tubes, we derived the tube conductance to be $C=\frac{1}{R}=\frac{\pi a^4}{8\mu l}$, see Eq.~\eqref{eq:ResHagenPoiseuille}. Now Kirchhoff's first law states that the flow rates at a network node sum to zero: 
\begin{align}\label{eq:KirchhoffCurrentLawFluid}
    q_i =\sum_{j\,\mathrm{n.n.}\,i} Q_{ij} = \sum_{j\,\mathrm{n.n.}\,i} C_{ij} (P_j - P_i)=0,
\end{align}
where $j$ runs over all the nearest neighbor nodes of $i$ and $q_i$ is the total fluid flow through node $i$, which is zero if flow is conserved but can be non-zero if flow is injected or extracted from the network at that node. The flows and pressures are coupled across the network as they may appear in multiple Kirchhoff conditions of type Eq.~\eqref{eq:KirchhoffCurrentLawFluid}. As a short-hand, we write Kirchhoff's first law for all nodes in a matrix form listing all node pressures $P_i$ and net node flows $q_i$ in $N$-dimensional vectors $\Vec{q}$ and $\Vec{P}$, respectively,
 \begin{align}
    \begin{pmatrix}
        \vdots \\
        q_i \\
        \vdots   
    \end{pmatrix} = 
    \begin{pmatrix}
         & \cdots & \\
         \vdots & C_{ij} & \vdots \\
          & \cdots & \\
    \end{pmatrix}
    \begin{pmatrix}
        \vdots \\
        P_j \\
        \vdots   
    \end{pmatrix},
    \label{eq:KirchhoffMatrix}
\end{align}
where $C_{ij}$ is the conductance of the tube connecting nodes $i$ and $j$ ($i\neq j$), and 
 $C_{ii} := -\sum_{j\,\mathrm{n.n.}\,i} C_{ij}$. Eq.~\eqref{eq:KirchhoffMatrix} can be written more succinctly as
 \begin{align}
     \Vec{q}=-\hat{L}\Vec{P}
 \end{align}
 by defining a matrix $\hat{L}$ with elements
 \begin{align}
     L_{ij}=-C_{ij}+\left(\sum_{k}C_{jk}\right)\delta_{ij},
 \end{align}
 where $\delta_{ij}$ is the Kronecker delta function. Note that $C_{ij}=C_{ji}$ is zero when nodes $i$ and $j$ are not connected. $\hat{L}$ is a symmetric positive semi-definite matrix, known as the weighted Laplacian of the network. 
 Given the conductance $C_{ij}$ of every tube within a network and the inflows and outflows $q_i$ at every network node, inverting the matrix $\hat{L}$ solves for the pressures $P_i$ at every network node. The flow rate across a single network tube $Q_{ij}$ then follows by equating $Q_{ij}=C_{ij}(P_i-P_j)$. However, it should be noted that directly inverting the matrix $\hat{L}$ is not possible, as the system is overdetermined. The reason is that only pressure gradients have a physical meaning, and not absolute pressures. Therefore, we need to set one nodal pressure as the reference pressure. A typical choice is to set the reference pressure to zero, which effectively reduces the linear system to a subsystem that is invertible. Altogether, flows throughout the network are fully determined by Kirchhoff's first law, flow boundary conditions, and network geometry specifying tube conductance. If a tube dilates or shrinks in diameter, its conductance is altered, which in turn changes the flow rates throughout the network. Therefore, local changes in tube conductance are globally coupled to flows in a flow network. Now biological flow networks may respond to local changes in flows and locally alter the tube geometry, initiating a global feedback cycle, as those local changes in tube geometry affect overall flows and thus change flows in other tubes, which may themselves adapt the tube geometry. This global coupling underlies much of the self-organizing capability of biological flow networks that we lay out in the following chapters. 

This short primer does not do justice to the beauty of fluid physics and the breadth of phenomena that all originate from the Navier-Stokes equation and the continuity equation. Wonderful books have been written that go beyond the short introduction here, for example, \cite{Acheson.1990,Batchelor_2000,truskey2009transport,Bruus.2007}. 

\section{Networks minimizing energy dissipation}
Viscous fluid flow is always associated with the dissipation of energy put into the system to generate the pressure gradients that drive the flow. How may a flow network adapt individual tube radii to minimize the dissipation of energy and what kind of network architectures arise thereby? We review both local and global adaptation rules and the impact of fluctuations in flows on the resulting network architecture. 
\subsection{The physiological principle of minimal work: Murray's law}
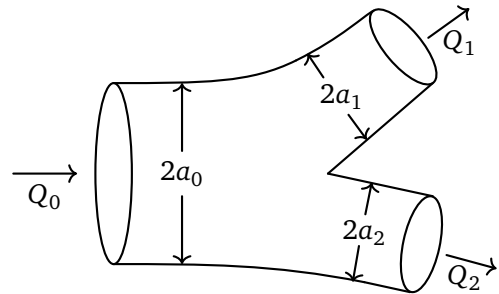
\begin{wrapfigure}{r}{0.5\textwidth}
\centering
\begin{tikzpicture}[line cap=round, line join=round, thick,scale=0.6]

\draw (1,5) .. controls (3,5) and (4,5) .. (5,5.5);
\draw (1,1) .. controls (3,1) and (4.5,1) .. (6.5,0.6);

\draw (5,5.5) .. controls (6,6) and (6.5,6.5) .. (6.7,6.6);

\draw (5.7,3) -- (8,5);

\draw (6.5,0.6)--(7.5,0.38);

\draw (5.7,3) -- (8,2.5);

\draw (14.7/2,11.6/2) ellipse [x radius=1.03cm, y radius=0.4cm, rotate=130];
\draw (15.5/2,2.88/2) ellipse [x radius=1.08cm, y radius=0.4cm, rotate=78];
\draw (1,3) ellipse [x radius=2cm, y radius=0.4cm, rotate=90];


\draw[->] (2.5,3.5) -- (2.5,5);
\node at (2.5,3) {$2a_0$};
\draw[->] (2.5,2.5) -- (2.5,1);

\node at (-0.5,2.5) {$Q_0$};
\draw[->] (-1.2,3) -- (0.2,3);

\draw[->] (5.6,5.125) -- (5.25,5.65);
\node at (6,4.7) {$2a_1$};
\draw[->] (6.15,4.3) -- (6.55,3.75);
\node at (8.6,5.9) {$Q_1$};
\draw[->] (7.9,6) -- (8.8,6.65);

\draw[->] (6.52,2.1) -- (6.66,2.8);
\node at (6.5,1.7) {$2a_2$};
\draw[->] (6.38,1.4) -- (6.245,0.61);
\node at (8.7,0.7) {$Q_2$};
\draw[->] (8.3,1.2) -- (9.4,0.9);

\end{tikzpicture}
\caption{\textbf{Murray's law exemplified at a three tube network node.} The tube radius $a_0$ of the tube with incoming flow $Q_0$ is related to the radii of the tubes $a_1$ and $a_2$ of outgoing flow $Q_1$ and $Q_2$, respectively, by Murray's law according to $a_0^3=a_1^3+a_2^3$, thereby minimizing both dissipation and metabolic cost.}
    \label{fig:Murray}
\end{wrapfigure}
In search of a quantitative description of physiology, Cecil D.~Murray in 1926 put forward the idea that physiology might be adapted to minimize work \cite{murray_physiological_1926}. In particular, the human vasculature was well quantified at that time to stand the test with a theoretical prediction. So Murray proposed that vascular networks may minimize energy dissipation at a fixed metabolic cost for maintaining the network. For a single tube of radius $a$ and length $l$ perfused by a fluid of viscosity $\mu$ at flow rate $Q$, the cost function $H$ is thus the sum of the flow energy dissipation and the tube volume: 
\begin{align}\label{eq:MurrayCost}
    H = \underbrace{\frac{8 \mu l Q^2}{\pi a^4}}_{\text{dissipation}} + \underbrace{\vphantom{\frac{8 \mu l Q^2}{\pi a^4}}b \pi a^2l}_{\text{metabolic cost}},
\end{align}
where $b$ is the metabolic cost per volume per unit time. In order to find the optimal tube radius that minimizes both dissipation and metabolic cost, the minimum of the cost function with respect to the radius $a$ is evaluated to yield a relation between the optimal tube radius and the flow rate:
\begin{align}\label{MurrayOptimalRadius}
    \pder[H]{a} &= -\frac{4 \cdot 8 \mu l Q^2 }{\pi a^5} + 2 b \pi l a = 0, \nonumber \\
    \implies Q&= \frac{\pi a^3}{4}\sqrt{\frac{b}{\mu}}.
\end{align}

 Testing such a prediction experimentally would require access not only to the tube geometry but also to the flow rate or flow velocity, which is typically very hard to obtain. However, Murray realized that the optimal radius prediction may translate to a purely geometric prediction when considering a network node. At a network node, Kirchhoff's first law states that to conserve fluid volume, flow rates need to sum to zero, see Eq.~\eqref{eq:KirchhoffCurrentLawFluid}. For a network node with one incoming flow, $Q_0$, and two outgoing flows, $Q_1$ and $Q_2$, see Fig. \ref{fig:Murray}, the incoming and outgoing flows are exactly balanced. Therefore, via the optimal radius prediction, Eq.~\eqref{MurrayOptimalRadius}, the radii of the incoming and outgoing tubes are related as follows:
\begin{align}
    Q_0 &= Q_1 + Q_2, \nonumber \\
    \implies \frac{\pi a_0^3}{4}\sqrt{\frac{b}{\mu}} &= \frac{\pi a_1^3}{4}\sqrt{\frac{b}{\mu}} + \frac{\pi a_2^3}{4}\sqrt{\frac{b}{\mu}}, \qquad \text{(from Eq.\eqref{MurrayOptimalRadius})}\nonumber \\
    \implies a_0^3 &= a_1^3 + a_2^3,    
\end{align}
 under the assumption that all tubes have the same metabolic cost per volume per unit time $b$ and the same fluid viscosity $\mu$. The cubic relation of the tube radii is known as Murray's law. An alternative derivation of Murray's law assumes that the shear stress at the tube wall  $\tau= \frac{4Q\mu}{\pi a^3}$ is constant at a network node instead of the optimization of tube radius, before again relating the flow rates across a network node by Kirchhoff's law.  Murray's law has since been found to agree with vascular networks across very different forms of life; from animals \cite{Sherman_1981,West.1997,Kassab.1995}, to plants \cite{West.1997,McCulloh.2003} and slime molds \cite{Akita.2016,Fricker.2017}. Yet, caution should be taken when testing Murray's law against experimental data, as Murray's law is non-linear in the scaling exponent, skewing fluctuations in tube radii, see \cite{Akita.2016}.  Therefore, it is desirable to predict the dynamics of tube adaptation during optimization. 

\subsection{Tube adaptation by shear stress feedback}
An equation of motion for tube radius dynamics can be derived by assuming that tube radii follow the gradient of the cost function towards its minimum in overdamped kinetics \cite{hu_adaptation_2013, ronellenfitsch_global_2016}. The gradient of the cost function, Eq.~\eqref{eq:MurrayCost}, herein acts as a force
\begin{align}\label{eq:ForceFromCostFunc}
    F = -\pder[H]{a} = \frac{4 \cdot 8 \mu l Q^2 }{\pi a^5} - 2 b \pi l a,
\end{align}
that is acting on the dynamics of an overdamped tube radius, where the friction $\beta$ dominates over the inertial term with mass $m$:
\begin{align}\label{eq:overdamped}
    F = \underbrace{\cancel{m\frac{d^2a}{dt^2}}}_{\text{inertia $\approx $ 0}}+\underbrace{\beta\frac{da}{dt}}_{\text{friction}}\approx \beta \frac{da}{dt},
\end{align}
such that the tube radius dynamics follow from Eq.~\eqref{eq:ForceFromCostFunc} and Eq.~\eqref{eq:overdamped},
\begin{align}
    \label{eq:TubeDyn}
    \frac{da}{dt}&=\frac{2\pi l a}{\mu \beta}\left(\frac{16\mu^2Q^2}{\pi^2a^6}-b\mu\right), \nonumber \\
    \implies \frac{da}{dt} &=\frac{2\pi a} {\tilde{\beta}}\left(\frac{\tau^2}{\tau_0^2} - 1\right), \qquad \qquad (\text{substituting $\tau=\frac{4Q\mu}{\pi a^3}$ from Eq.~\eqref{eq:WallStress}})
\end{align}
where $\tau_0=\sqrt{b\mu}$ is an effective optimal shear stress and $\tilde{\beta}=\beta/lb$ is the time scale of tube adaptation. Thus, the tube adaptation dynamics is driven by the shear stress adapting to an optimal shear stress $\tau_0$. This implies that tubes where the shear stress exceeds the optimal shear stress $\tau_0$ grow in radius, while those at lower shear stress shrink. Note that the optimal shear stress $\tau_0=\sqrt{b \mu}$ increases when the metabolic cost of building the tubes is higher. Therefore, the more expensive the metabolic cost, the less tube diameter growth and the less hierarchy in tube radii across a network. 

The ease of an optimization principle comes to light here as tube adaptation dynamics directly follow in a few lines from the cost function, and the impact of the balance of costs on adaptation dynamics is readily read off. Yet, at the same time, an optimization principle occludes the physical meaning of parameters such as the optimal shear stress $\tau_0$ and the adaptation time scale $\tilde{\beta}$, and how changes in physical parameters would affect them. Such insight can only be provided when deriving dynamics directly from force balance, yet here it is potentially limited to the specific system under consideration.
\subsection{Tube adaptation dynamics from force balance}
For tube adaptation dynamics, both the top-down approach minimizing dissipation at finite metabolic cost, and a bottom-up approach deriving tube dynamics from force balance at the tube wall yield the same dynamics of Eq.~\eqref{eq:TubeDyn}. Motivated by direct observation of tube radius adaptation dynamics in slime molds, we consider a tube with a thin, visco-elastic tube wall \cite{marbach_vascular_2023, marbach_vein_2023}. Coarse-graining over the short-time tube dynamics in slime molds \cite{marbach_vascular_2023, marbach_vein_2023}, here we directly turn to the long-time tube radii adaption.  The long-time viscous yielding of the tube wall of elastic viscosity $\eta$, Poisson ratio $\nu$, and wall thickness $e$ balances the stresses directed radially on the tube wall:
\begin{align}\label{eq:TubeForceBalance}
    \frac{\eta}{(1-\nu^2)}\frac{1}{a}\frac{da}{dt}= \frac{a}{e}\left[(p-p_{\text{ext}}) + \sigma_{\text{active}}+\frac{\mu}{t_{\text{wall}}}\frac{\tau^2}{\tau_c^2} \right],
\end{align}
where $(p-p_{\text{ext}})$ is the hydrostatic pressure difference between the inside and outside of the tube, $\sigma_{\text{active}}$ is an active stress in the tube wall that contracts the tube and $\frac{\mu\tau^2}{t_{\text{wall}}\tau_c^2} $ is a normal stress response due to the flow shear stress acting tangentially to the tube wall \cite{Vahabi.2018, Conti.2009, Holzapfel.2014,Janmey.2007} relative to a characteristic shear stress $\tau_c$ and the wall's characteristic response time $t_{\text{wall}}$. The normal stress response to a tangential shear is a mechanical property of semi-flexible polymer gels, in which the characteristic shear stress $\tau_c$ and characteristic response time $t_{\text{wall}}$ depend on the mechanical properties of the gel, such as porosity and cross-linking. The endothelial cells lining the blood vasculature also mount a normal stress in response to tangential shear stress \cite{Livne.2014,Landau:2018}. Now, identifying the optimal shear stress as $-\mu\tau^2_0/t_{\text{wall}}\tau^2_c=(p-p_{\text{ext}})+\sigma_{\text{active}}$ and the adaptation time scale as $\tilde{\beta}=2\pi e \eta \tau_c^2 t_{\text{wall}}/\tau_0^2a\mu(1-\nu^2)$, we recover the local tube radii adaptation dynamics, which follows from the minimization of energy dissipation at a finite metabolic cost Eq.~\eqref{eq:TubeDyn}.

\begin{figure}[ht]
\begin{center}
\includegraphics[width=\textwidth]{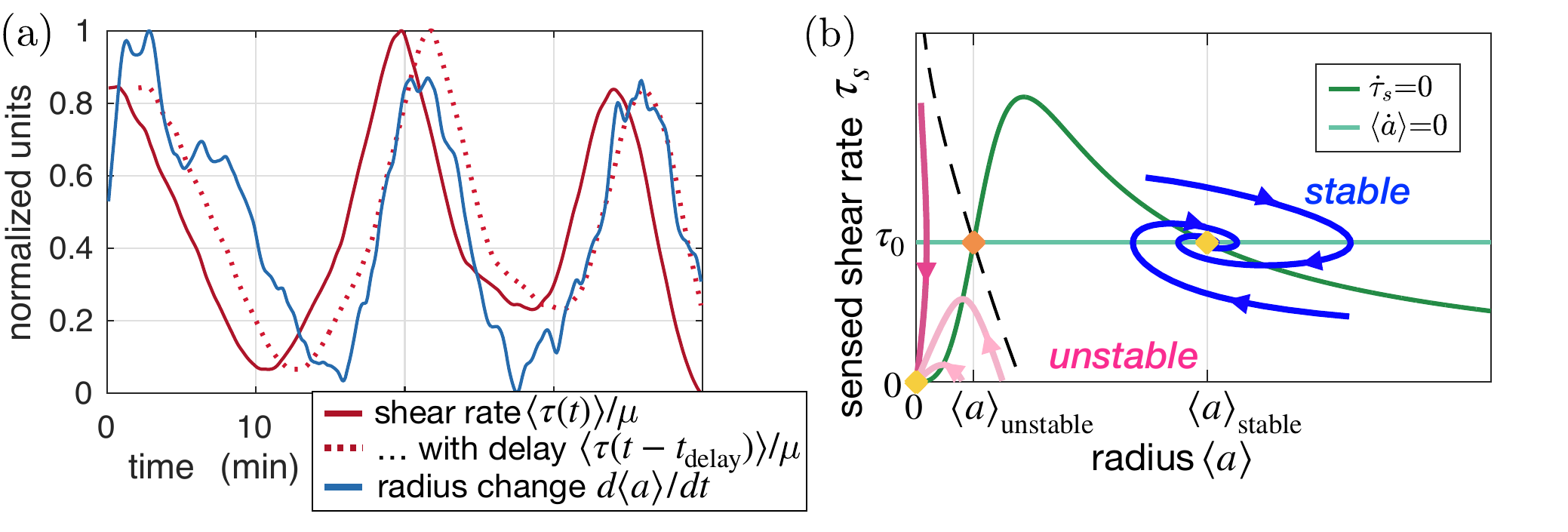}
\caption{ \textbf{Shear stress induces tube radius adaptation with a time delay in the slime mold \textit{Physarum polycephalum}.} (a) Time-averaged  shear rate $\langle\tau\rangle/\mu$ dynamics preceed time-averaged tube radius dynamics $d\langle a\rangle /dt$ with a time delay $t_{\text{delay}}$. (b) Deriving the local shear rate within a flow network, the non-linear dynamics of sensed shear rate $\tau_s$ and tube radius  $\langle a\rangle$ are mapped out in a phase portrait. Two stable fixed points (light yellow) constrain tube trajectories to circle in spirals (blue) or shrink away (pink). Reproduced from (Marbach et.~al.~(2023) \cite{marbach_vein_2023}).} 
\label{fig:Marbach}
\end{center}
\end{figure}

Experimental recordings of tube radius dynamics and flow shear rate, i.e.~shear stress normalized by fluid viscosity, in the slime mold  \textit{Physarum polycephalum} reveal that tube radii do not respond instantly to flow shear stress, but rather with a time delay \cite{marbach_vein_2023}, see Fig.~\ref{fig:Marbach}. This time delay results in two possible tube radii dynamics, either a continuous decrease in shear stress and tube radius until the tube gets vanished, or oscillating spirals of shear stress and tube radius about a fixed point of finite shear stress and tube radius. Here, the stable fixed point corresponds to the steady-state tube radius predicted by Murray's law, Eq.~\eqref{MurrayOptimalRadius}. Therefore, Murray's law captures well the steady state of biological flow networks. It is noteworthy that the small time-delay has a conceptional impact. Due to the delay in response the tube can locally contrast current and past flow rate, which enables tubular flow networks to perform contrastive learning tasks \cite{Falk.2025}. Beyond the scope of optimizing network architecture, flow shear force adaptation solves complex tasks such as the traveling salesman problem \cite{Tero_2010}. Shear force adaptation by analytical form, further, resembles Hebbian adaptation dynamics of neurons and embeds adaptive flow networks as an utterly widespread building block of life that may exhibit physical learning dynamics \cite{Stern.2023}.

\subsection{Generalization of metabolic cost scaling}
Originally, Murray's cost function assumed the metabolic cost to be proportional to the volume of the tube, i.e.~to grow quadratically in tube radius. However, the metabolic cost may also be dominated by the cost for the tube wall only, thus scaling linearly in tube radius. It therefore makes sense to introduce a general scaling variable in the metabolic cost \cite{Bohn.2007}. The choice is to re-define the metabolic cost in terms of the tube conductance $C=\frac{\pi a^4}{8\mu l}$. For Poiseuille flow, see Table \ref{tab:ElectroHydro}, recasting Murray's original metabolic cost  in Eq.~\eqref{eq:MurrayCost}) results in
\begin{align}\label{eq:ConductanceVol}
    b\pi a^2 l = b \sqrt{8 \pi l^3 \mu} C^{1/2}.
\end{align} 
Introducing $\gamma$ as a parameter for the scaling exponent of the metabolic cost, the generalized cost function $H_\gamma$ for an entire network is therefore obtained by summing over every tube $jk$ connecting nodes $j$ and $k$ in the network \cite{corson_fluctuations_2010,katifori_damage_2010}:
\begin{align}\label{eq:GenCostFunc}
    H_\gamma = \sum_{jk}\left(\frac{Q_{jk}^2}{C_{jk}}+\xi b C_{jk}^\gamma\right),
\end{align}
absorbing other constants into $\xi$. In this context, a metabolic cost dominated by the tube wall would correspond to a scaling with $C^{1/4}$. In leaf venation networks, photosynthetic activity scaling with the leaf area perfused by the venation may suggest scaling exponents between $1/2$ and $1$ \cite{hu_adaptation_2013}. This closed expression now describes the cost function of a flow network only as function of the flow rates $Q_{ij}$ and the conductance $C_{ij}$ of each tube. It therefore makes sense to also recast tube adaptation dynamics in terms of changes in tube conductance \cite{hu_adaptation_2013}, 
\begin{align}\label{eq:GenTubeDyn}
    \frac{dC_{ij}}{dt} =\frac{C_{ij}}{\beta'}\left( \frac{Q_{ij}^2}{C_{ij}^{\gamma+1}}-\tau_0^2 \right),
\end{align}
which recovers Eq.~\eqref{eq:TubeDyn} for $\gamma=1/2$, i.e.~by identifying $Q_{ij}^2/C_{ij}^{\gamma+1}$ with the wall shear stress. Note that for simplicity, we throughout assumed networks to consist of tubes of equal length $l$, only varying in tube radius $a$. To take into account variation in tube lengths, a tube's individual conductance needs to be normalized by the corresponding tube length, see \cite{hu_adaptation_2013,ronellenfitsch_global_2016,bhattacharyya_memory_2022}.

The cost function $H_\gamma$, Eq.~\eqref{eq:GenCostFunc}, and the local conductance adaptation dynamics, Eq.~\eqref{eq:GenTubeDyn}, represent two entirely different optimization schemes. Minimizing the cost function $H_{\gamma}$ is global, where knowledge of the entire network state is taken into account in the search for the global optimum. The local adaptation dynamics, however, only take into account local information of the conductance and the flow rate at the given tube that is adapting. Such local adaptation dynamics are, therefore, more likely to be at play in a biological system. Yet, we do not expect local dynamics to find the global minimum, but to more likely get stuck in local minima. 
\subsection{Optimizing network architecture for minimal energy dissipation}
To assess which network architecture minimizes energy dissipation at a finite metabolic cost, we first review the task of minimizing the cost function globally, Eq.~\eqref{eq:GenCostFunc} \cite{corson_fluctuations_2010,katifori_damage_2010,Bohn.2007,durand_structure_2007}. The analytical form of the cost functions allows one to derive a closed expression for the iterative adaptation of all network tubes, opening the route for efficient numerical minimization and analytical consideration. To derive the global minimization scheme, the metabolic cost per volume and time, $b$, is treated as a Lagrange multiplier which constrains the metabolic cost to a fixed value $\sum_{ij} C_{ij}^\gamma=K^\gamma$. Equating the extremum of the cost function $H_{\gamma}$, Eq.~\eqref{eq:GenCostFunc}, with respect to the tube conductance $C_{ij}$ solves for the Lagrange multiplier:
\begin{align}\label{eq:GamLam}
    \pder[H_\gamma]{C_{ij}} &= -\frac{Q_{ij}^2}{C_{ij}^2}-b\xi\gamma C_{ij}^{\gamma-1}=0, \nonumber\\
    \implies -b\xi\gamma &= \frac{Q_{ij}^2}{ C_{ij}^{\gamma+1}}.
\end{align}
The Lagrange multiplier now also constrains the fixed metabolic cost $\sum_{ij} C_{ij}^\gamma=K^\gamma$ by rearranging,
\begin{align}
    \implies (Q_{jk}^2)^{\frac{\gamma}{\gamma+1}}&=(-b\xi\gamma)^{\frac{\gamma}{\gamma+1}} C_{jk}^{\gamma} \nonumber\\
    \implies \sum_{jk}(Q_{jk}^2)^{\frac{\gamma}{\gamma+1}}&=(-b\xi\gamma)^{\frac{\gamma}{\gamma+1}} \sum_{jk}C_{jk}^{\gamma} \nonumber \\
    \implies \sum_{jk}(Q_{jk}^2)^{\frac{\gamma}{\gamma+1}}&= (-b\xi\gamma)^{\frac{\gamma}{\gamma+1}} K^\gamma.
\end{align}
Substituting $(-b\xi\gamma)$ from Eq.~\eqref{eq:GamLam} results in a direct expression for the optimal conductance at a given flow pattern throughout the network:
\begin{align}\label{eq:OptConductance}
    C_{mn}^\gamma &= K^\gamma\frac{(Q_{mn}^2)^{\frac{\gamma}{\gamma+1}}}{\sum_{jk}(Q_{jk}^2)^{\frac{\gamma}{\gamma+1}}}, \nonumber \\
    \implies C_{mn} &= K\frac{(Q_{mn}^2)^{\frac{1}{\gamma+1}}}{\left(\sum_{ij}(Q_{ij}^2)^{\frac{\gamma}{\gamma+1}}\right)^{1/\gamma}}.
\end{align}
With this expression for the optimal conductance, the global optimization can effectively proceed as follows. A network is drawn at a given topology and tube conductance. In- and out-flows at nodes are specified, and together with the tube conductance, they determine the flow rates in individual tubes. Then, iteratively, tube conductances are updated according to Eq.~\eqref{eq:OptConductance} and the flow rates are re-calculated. 

Flow patterns for minimal dissipation networks are often motivated by leaves where a single inlet at the leaf stalk is servicing the entire leaf venation network with water which evaporates throughout the entire leaf \cite{Roth-Nebelsick_2001}. During leaf venation formation, the opposite transport of the plant hormone auxin produced throughout the entire leaf and transported to the plant stem lays out the venation pattern \cite{Dimitrov_2006}. The leaf blade is thus represented by a regular grid with a single outflow at a node in the corner of the grid, which is serviced by equal inflows at all other network nodes. With such flow boundary conditions, the minimal dissipation network at a finite metabolic cost is a tree with tubes that have smaller and smaller radii the further they are from the outlet for $\gamma <1$, and for $\gamma >1$, it is a finely reticulated loopy network lacking any hierarchy in tube radius \cite{Bohn.2007,Banavar.2000, durand_structure_2007}. The scaling exponent $\gamma$ of the metabolic cost here determines the cost of large radii. The larger $\gamma$ is, the more expensive the cost of large radii. Thus, hierarchy in tube conductance becomes suppressed at large $\gamma$. Numerical optimization of the network architecture for $\gamma<1$ typically leads to a local minimum; the global minimum may only be found on particularly searching the space of tree topologies \cite{Bohn.2007}. Thus, the energy landscape of optimal tree architectures is rugged, akin to disordered systems. Alternatively, coupling cost function minimization to the growth of the tissue underlying the tubular network can drastically improve the search for the global optimum \cite{ronellenfitsch_global_2016}. In combination with the adaptation dynamics on network nodes, global optima considerably resemble real leaf venation networks \cite{Alonso.202411r}.

\begin{figure}[t]
\begin{center}
\includegraphics[width=\textwidth]{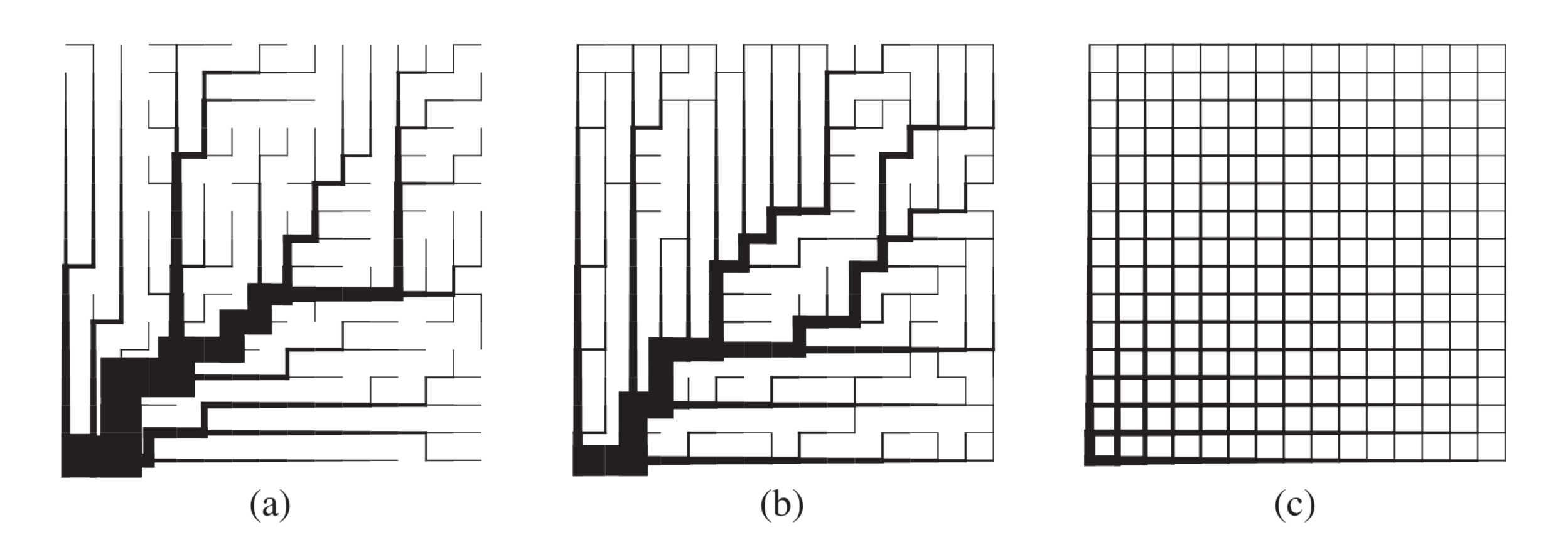}
\caption{\textbf{Minimal dissipation networks with fluctuating outflows transition from tree to loopy networks as metabolic cost becomes more and more expensive for large tube radii.} In each network, the inflow is at the lower left corner and all other nodes are outlets with uncorrelated fluctuating flows. The radius of each tube is proportional to the square root of its conductance. (a) Tree-like network ($\gamma = 0.25$). (b) Hierarchical network with loops ($\gamma=0.75$). (c) Network with many loops and no hierarchical organization ($\gamma=1.25$). Reproduced from Corson (2010) \cite{corson_fluctuations_2010}.} 
\label{fig:CorsonNetwork}
\end{center}
\end{figure}

\begin{figure}[t]
\begin{center}
\includegraphics[width=0.8\textwidth]{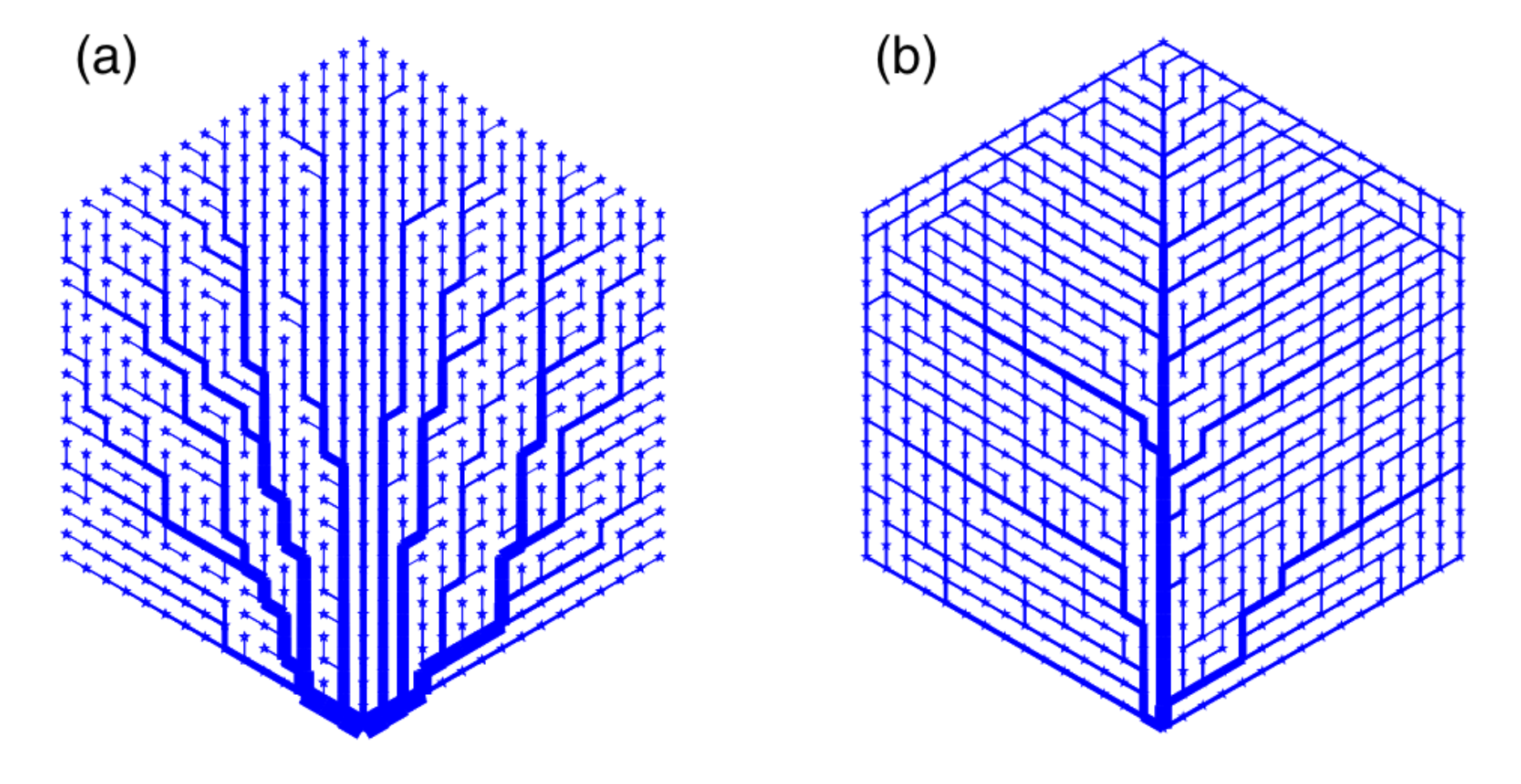}
\caption{\textbf{Fluctuations introduce loops also with local adaptation dynamics.} In each network, the inflow is at the lowest node and distributed throughout all nodes serving as outlets in the remaining network, $\gamma=1/2$. (a) Fixed and equal outflow at all nodes results in a tree architecture   (b) Outlet fluctuate as the tube radii are adapting, resulting in loops interconnecting the tree architecture. Reproduced from Hu and Cai (2013) \cite{hu_adaptation_2013}.} 
\label{fig:HuNetwork}
\end{center}
\end{figure}

If inflows fluctuate either in time \cite{corson_fluctuations_2010} or by moving throughout the network \cite{katifori_damage_2010}, loops appear in the network already at $\gamma <1$, see Fig.~\ref{fig:CorsonNetwork}. The alternate routes for transport provided by the loops ensure that also large local flows arising from fluctuations can be efficiently transported, thereby deviating from the otherwise tree-like optimal network architecture. Loops also provide robustness against damage \cite{Bebber_2007, katifori_damage_2010}.
Notably, the core result that networks are tree-like if the cost for very large radii close to the outlet is not exorbitant, i.e.~$\gamma<1$, and loopy without radii hierarchy otherwise, and that fluctuations in flows introduce loops also at $\gamma<1$, persists even if we turn to local adaptation dynamics of Eq.~\eqref{eq:GenTubeDyn}, see Fig.~\ref{fig:HuNetwork}. That local and global dynamics result in qualitatively the same network architecture is non-trivial per se, as the local adaptation dynamics do not constrain the metabolic cost to a fixed value anymore. Instead, a tree-like network has a high dissipation because of the high flow rate through its big tubes, but saves on the metabolic cost by vanishing small tubes. On the contrary, reticulate networks show low dissipation as flows are distributed and small, but need to invest more in metabolic cost, for maintaining all loopy connections \cite{ronellenfitsch_phenotypes_2019,Schick:2024}.
However, fluctuations per se do not guarantee loops at $\gamma<1$. The more fluctuations are spatially correlated, the more loops are suppressed in the network \cite{ronellenfitsch_phenotypes_2019}. The lack of loops at large spatial correlations follows as we expect to recover the optimal tree architecture at infinite spatial correlations of fluctuations, which represents a state of no fluctuations at all. 
\section{Memory formation in adaptive flow networks}
Optimization of flow networks for minimal dissipation at a finite metabolic cost revealed a rugged energy landscape of disordered systems. Since disordered systems may retain memories of the past \cite{Murugan.2015,Keim.2011,Keim.2019,Rocks.2019ehd}, flow networks adapting to minimize dissipation may also imprint past flow patterns. 
\subsection{Adaptive flow networks retain memory of past inflow}
\begin{figure}[ht]
\begin{center}
\includegraphics[width=\textwidth]{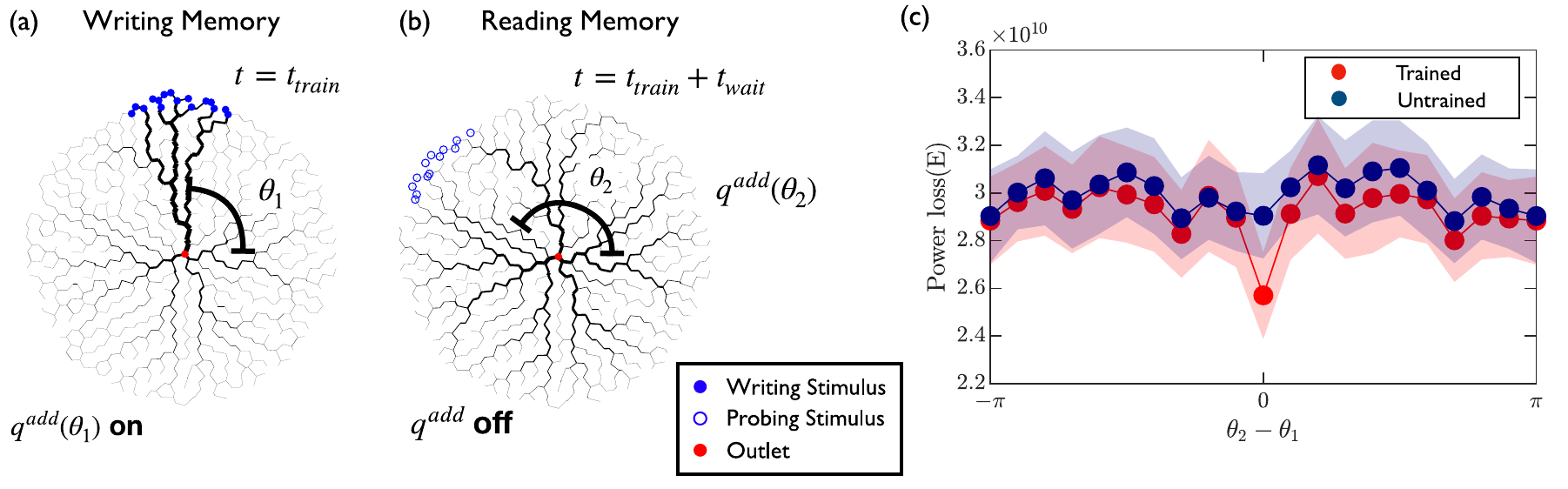}
\caption{\textbf{Networks with $\gamma=1/2$ undergo memory formation and retrieval.} (a) Network trained by adapting at  additional inflow $q^{\text{add}}$ , equally distributed over outer nodes around $\theta_1$ (blue filled), for training duration $t_{\text{train}}$. Centered outlet node depicted in red. (b) Network subsequently adapted without additional inflow over waiting time $t_{\text{wait}}$. Memory is subsequently probed by applying probing inflow at angle $\theta_2$ (empty blue). (c) Power loss $E=H_\gamma$ over 200 independent simulations versus $(\theta_2 - \theta_1)$ for varying $\theta_2$ for trained dataset (red) and for untrained data set (blue). Reproduced from Bhattacharyya et. al. (2022) \cite{bhattacharyya_memory_2022}.} 
\label{fig:BhattPRLModel}
\end{center}
\end{figure}
To test memory formation of past flow patterns, a disk-shaped network is considered, where the central node is the sole outlet for the inflows at all other nodes throughout the network. As before, inflows fluctuate by stochastically flipping between an open and closed state with a time-average inflow of $q^{(0)}$ per node, see Fig.~\ref{fig:BhattPRLModel}. The network adapts at every numerical time step as all tube conductances are updated to minimize dissipation at a constant metabolic cost $K^\gamma=\sum_{ij}C_{ij}^\gamma$, as derived in the previous chapter in Eq.~\eqref{eq:OptConductance}. Initially, metabolic cost is set to scale as $\gamma=1/2$ to focus on loopy tree-like optimal networks. Now, at network initialization, an additional inflow of magnitude $q^{\text{add}}(\theta_1)$ is applied at few nodes at the network boundary, positioned at a chosen radial angle $\theta_1$ in the disk-shaped network, see Fig.~\ref{fig:BhattPRLModel}(a), for a duration $t_{\text{train}}$. The network evolves into a loppy, tree-like architecture with a particularly eminent tree architecture towards the additional inflow. Then, the additional inflow is removed and the network adapts for a duration $t_{\text{wait}}$ with the usual stochastic open-close switching of equal inflows at all network nodes. The network returns to a seemingly isotropic architecture, see Fig.~\ref{fig:BhattPRLModel}(b). However, assessing the retained memory by probing the network at all angles $\theta_2$ with a probing inflow and quantifying the cost function $H_\gamma$, one observes that the network minimizes the cost function precisely at the angle $\theta_1$, where the training stimulus was provided, see Fig.~\ref{fig:BhattPRLModel}(c). Thus, memory of past flow is retained in the network's energy landscape.
\subsection{Mechanism of memory formation}
\begin{figure}[ht]
\begin{center}
\includegraphics[width=\textwidth]{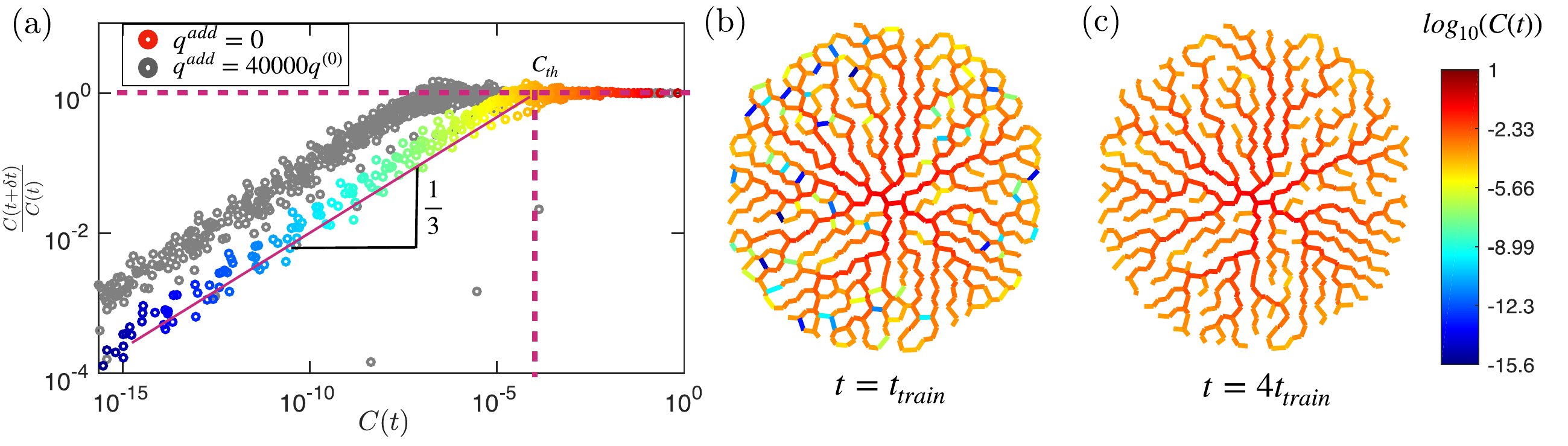}
\caption{\textbf{Memory formation due to disappearance of low conductance tubes.} (a) Ratio of conductance of two subsequent iterations versus preceding conductance during adaptation for $3t_{\text{train}}$ iterations after training phase of duration $t_{\text{train}}$ ended. Above threshold conductance $C_{\text{th}}$ (vertical red dashed line), conductances fluctuate around $|C(t+\delta t)/C(t)|=1$ (horizontal red dashed line). Low conductance tubes follow a power law with exponent $1/3$ (red line). Only threshold conductance $C_{\text{th}}$ is stimulus strength specific; compare gray $(q^{\text{add}}=40000q^{(0)})$ and color $(q^{\text{add}}=0)$. (b) A network adapted for $t_{\text{train}}$ and (c) the same network trained for longer duration, $4t_{\text{train}}$. Links with conductance smaller than threshold $C_{\text{th}}$ disappear in (c) Reproduced from Bhattacharyya et. al. (2022) \cite{bhattacharyya_memory_2022}.} 
\label{fig:BhattPRLMemoryStorage}
\end{center}
\end{figure}
\begin{figure}
\begin{center}
\includegraphics[width=\textwidth]{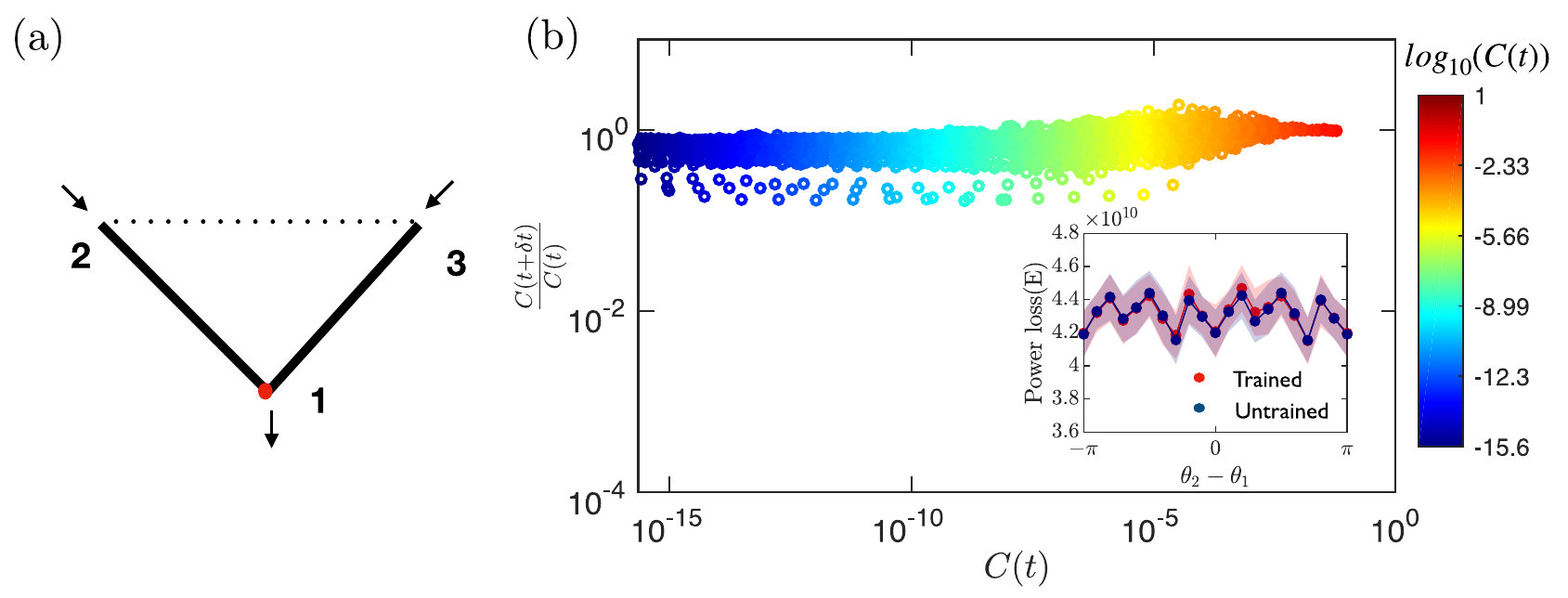}
\caption{\textbf{Memory formation limited to networks of $\gamma<1$ metabolic cost scaling.} (a) Simple network of two inlets, node 2 and 3, and a common outlet at node 1 (red dot) used to derive the adaptation dynamics. The network adapts to attain an architecture that is close to the optimal tree architecture, with a tube of vanishingly small conductance (dotted line) interconnecting the inlets. (b) Memory formation is lost when the metabolic cost scales with $\gamma=1$. In this case, all tubes persist and the power loss $E=H_\gamma$ in trained and untrained network is  identical (inset). Reproduced from Bhattacharyya et. al. (2022) \cite{bhattacharyya_memory_2022}.} 
\label{fig:BhattPRLSimple&Gamma}
\end{center}
\end{figure}

Quantifying network adaptation dynamics over subsequent numerical times steps $\delta t$ reveals that tubes with low conductances continuously decrease in their conductance, see Fig.~\ref{fig:BhattPRLMemoryStorage}. Therefore, low conductance tubes break the ergodicity of the system, cementing memories into the network architecture by shrinking away tubes that cannot regrow. Specifically, tubes with conductances above a certain threshold value $C_{\text{th}}$ fluctuate minimally while tubes of conductance lower than the threshold value shrink in conductance with a power law behavior:
\begin{align}\label{eq:ConductanceEvolution}  
    \frac{C(t+\delta t)}{C(t)} \approx \begin{cases}
        \left[ \frac{C(t)}{C_{\text{th}}} \right]^\omega \qquad &C(t)<C_{\text{th}}\\
        1 \qquad &C(t)\geq C_{\text{th}}.
    \end{cases}
\end{align}
This numerical observation can be analytically derived by working through the calculation of fluid flows in networks Eq.~\eqref{eq:KirchhoffMatrix} and the adaptation dynamics Eq.~\eqref{eq:OptConductance} in a network consisting of only three tubes, as outlined below. The analytical derivation predicts the power law decay of the low conductance tubes with an exponent $\omega=1/3$, in agreement with numerical observations, see Fig.~\ref{fig:BhattPRLMemoryStorage}(a). To arrive at this analytical result, consider a simple network with three nodes where node 1 is the outlet and the other nodes act as inlets, see Fig.~\ref{fig:BhattPRLSimple&Gamma}(a). The inlets are connected to each other and the outlet. Since  the optimal network architecture is a tree, we expect this small network to adapt into an architecture where the two tubes connecting the inlets and the outlet $C_{12}$ and $C_{13}$ are much larger than the tube connecting both inlets $C_{23}$. Therefore, deriving the tube adaptation dynamics of the inter-inlet tube $C_{23}$ should inform us about the dynamics of low conductance tubes. Having in mind that the inter-inlet tube adapts according to Eq.~\eqref{eq:OptConductance}
\begin{align}\label{eq:VNetConductance}
    C_{23}(t+\delta t)=Q_{23}(t)^{2/\gamma+1} \cdot \frac{K}{\left(\sum_{i, j} Q_{i j}(t)^{\gamma / \gamma+1} \right)^{(1/\gamma)}},
\end{align}
the flow rate throughout the network is required. Denoting the net inflow at node $i$ by $q_i$ and pressure at node $i$ by $P_i$,the following set of equations using Kirchhoff's laws from Eq.~\eqref{eq:KirchhoffMatrix} specifies the node pressures,
\begin{align*}
   \left[\begin{array}{l}
    q_1 \\
    q_2 \\
    q_3
    \end{array}\right]=\left[\begin{array}{ccc}
    C_{13}+C_{12} & -C_{12} & -C_{13} \\
    -C_{12} & C_{12}+C_{23} & -C_{23} \\
    -C_{13} & -C_{23} & C_{13}+C_{23}
    \end{array}\right]\left[\begin{array}{c}
    P_1 \\
    P_2 \\
    P_3
    \end{array}\right] . 
\end{align*}
Since node 1 is the outlet, its pressure is set to zero, $P_1=0$, to obtain
\begin{align}
\left[\begin{array}{cc}
C_{12}+C_{23} & -C_{23} \\
-C_{23} & C_{13}+C_{23}
\end{array}\right]^{-1}\left[\begin{array}{l}
q_2 \\
q_3
\end{array}\right]=\left[\begin{array}{c}
P_2 \\
P_3
\end{array}\right],   
\end{align}
resulting in a pressure drop between the two inlets equating to:
\begin{align}
  P_2-P_3=\frac{C_{13} q_2-C_{12} q_3}{C_{13}\left(C_{12}+C_{23}\right)+C_{12} C_{23}}.  
\end{align}
Therefore, the flow rate $Q_{23}$ between the two inlets is given by
\begin{align}\label{eq:VNetFlowRate}
    Q_{23}(t)=C_{23}\left(P_2-P_3\right)=C_{23} \frac{C_{13} q_2-C_{12} q_3}{C_{13}\left(C_{12}+C_{23}\right)+C_{12} C_{23}}\approx C_{23} \frac{C_{13} q_2-C_{12} q_3}{C_{13}C_{12}}.
\end{align}
Now, taking into account that the conductance of the inter-inlet tube $C_{23}$ is much smaller than the conductances of the tubes connecting to the outlet in the optimal tree architecture, $C_{12},C_{13}\gg C_{23}$, the denominator specifying the flow rate is governed by the product $C_{12}C_{13}$. As this flow in the inter-inlet tube is small compared to the flow rate in the tubes connecting to the outlets, the large flows are governed by the inflows at the upstream nodes, respectively, i.e.~$Q_{12}=q_2$ and $Q_{13}=q_3$. Note that flows determine tube conductances $C_{12}$ and $C_{13}$ through Eq.~\eqref{eq:OptConductance}. Therefore, the flow in the inter-inlet tube at a fluctuation in inflows say at $q_2(t)=q_2+\delta q$ with $q_3(t)=q_3$ is given by 
\begin{align}\label{eq:VNetFlowFinal}
  Q_{23}(t)=C_{23}(t) \frac{\left(q_3^{2 / \gamma+1}\left(q_2+\delta q\right)-q_2^{2 / \gamma+1}q_3\right)}{\left(q_2 q_3\right)^{2 / 
\gamma +1}} \frac{\left(q_2^{\gamma / \gamma+1} +q_3^{\gamma / \gamma+1} \right)^{1/\gamma}}{K}.   
\end{align}
Inserting this flow rate, Eq.~\eqref{eq:VNetFlowFinal}, into adaptation for the inter-inlet tube, Eq.~\eqref{eq:VNetConductance}, the change in conductance $C(t+\delta t)$ relative to its previous conductance $C(t)$ is,
\begin{align}\label{eq:CondThreshGeneral}
\frac{C_{23}(t+\delta t)}{C_{23}(t)} & =\left[\frac{1}{C_{t h}}\right]^{\frac{1 -\gamma}{1+\gamma} } C_{23}(t)^{\frac{1 -\gamma}{1+\gamma}},
\end{align}
where $C_{\text{th}}$ is a function of the flow rates, given by \cite{bhattacharyya_memory_2022}
\begin{align}\label{eq:ThresholdConductance}
    C_{\text{th}}=K \frac{(q_2q_3)^{\frac{4}{1-\gamma^2}} \Big( (q_2+\delta q)^{\frac{2\gamma}{1+\gamma}}+q_3^{\frac{2\gamma}{1+\gamma}}\Big)^{\frac{1+\gamma}{\gamma(1-\gamma)}}}{\Big(q_3^{\frac{2}{1+\gamma}}(q_2+\delta q)-q_2^{\frac{2}{1+\gamma}}q_3\Big)^{\frac{2}{1-\gamma}}\Big(q_2^{\frac{2\gamma}{1+\gamma}}+q_3^{\frac{2\gamma}{1+\gamma}} \Big)^{\frac{2}{\gamma(1-\gamma)}}}\;.
\end{align}
Inserting $\gamma=1/2$ into Eq.~\eqref{eq:CondThreshGeneral}, the numerically observed power law of $1/3$ is uncovered by which the conductances lower than the threshold conductance $C_{\text{th}}$ decay to zero. But there is an even more powerful result to be read off from Eq.~\eqref{eq:CondThreshGeneral}, namely, that for $\gamma\geq 1$, tube conductances do not decay at all. Therefore, the short analytical derivation predicts that networks with  $\gamma\geq 1$ cannot store memories, as ergodicity cannot be broken. Indeed, no memories were found on numerically probing networks of $\gamma=1$, see Fig.~\ref{fig:BhattPRLSimple&Gamma} (b). The large cost of large conductances, i.e.~large tube radii, at $\gamma\geq 1$, prevents hierarchy in tube radii from forming. Yet, for $\gamma<1$,  memories of past flows persist, and even multiple flow patterns can be remembered at the same time \cite{bhattacharyya_memory_2023}. The only limitation is that over time, small conductance tubes vanish, thereby effectively aging the network and reducing its ability to store new memories. Note that not only flow shear stress adaptation entails memory formation in adaptive flow networks. Chemicals released at a specific location in a flow network and dilating tube radii where they spread by flow, also imprints lasting memories \cite{Kramar.2021}.


\section{Optimizing for distribution and homogeneous supply in flow networks}
The tree-like or loopy flow network architectures predicted for minimizing dissipation at a finite metabolic cost do reminisce of leaf venation patterns with a hierarchy in vein radii and small loopy connections between the big veins. And yet, the success of minimizing dissipation at a finite metabolic cost is tied to the flow boundary conditions, namely, a single inlet and distributed outlets. On considering a single inlet and a single outlet, say, to describe the vasculature pervading an organ, the optimal flow network would simply be a single, chunky tube connecting the inlet and the outlet along the shortest path \cite{Hacking_1996}. This so-called ``shunt'' between the inlet and the outlet does not resemble our biological observations of a finely reticulated network. The minimal dissipation shunt would efficiently transport resources from the inlet to the outlet, but the shunt would not transport any resources to the tissue or collect products from the tissue.  In search for additional or alternative cost functions that flow networks optimize for, here we revisit two exemplary concepts - the optimization for distribution and homogeneous supply. In a yet alternative direction to transport and minimal dissipation, note that pressure has also been put forward as a local stimulus \cite{Pries_1998}.
\subsection{Optimizing transport in distributed production networks}
\begin{figure}[ht]
\begin{center}
\includegraphics[width=0.8\textwidth]{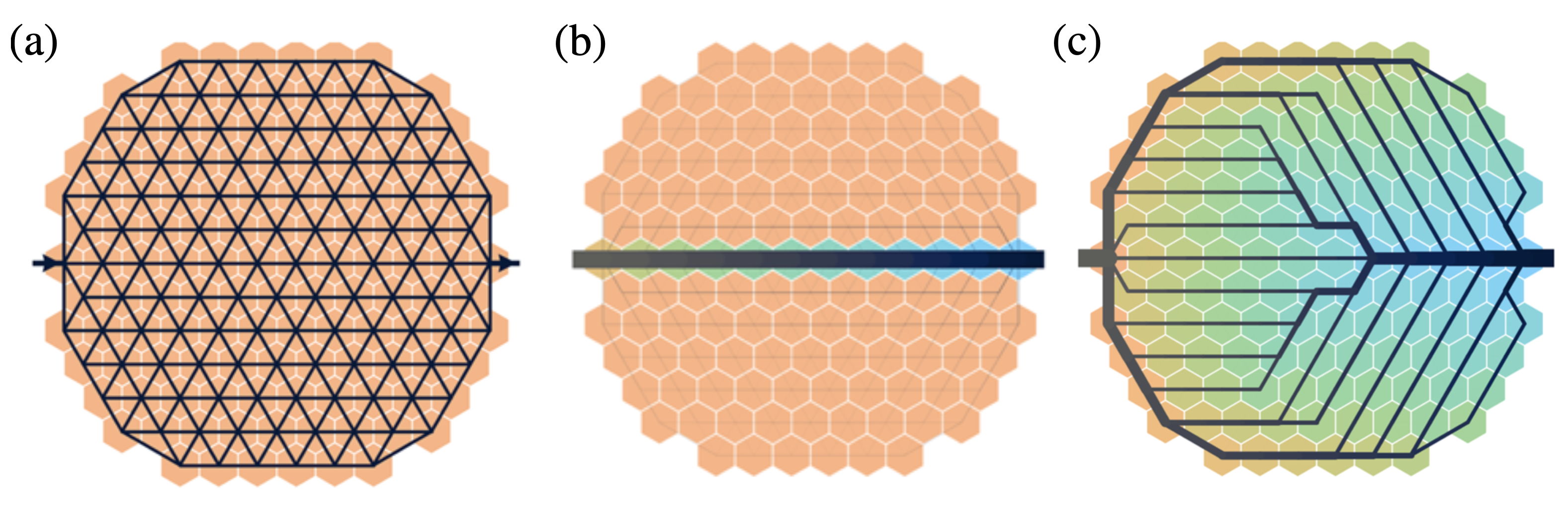}
\caption{\textbf{Optimal network architectures under single inlet, single outlet boundary conditions vastly differ.} (a) Initial network geometry with a single inlet (left) and a single outlet (right). (b) Minimizing dissipation at finite metabolic cost predicts a single ``shunt'' connecting the inlet and the outlet along the shortest path. (c) Minimizing average travel time from all network nodes to outlet generates a periphery - center network architecture. Cell colors indicate the average time to the outlet. Flow lines are colored by pressure, their thickness indicating the absolute value of the flow rate. Reproduced from Kirkegaard and Sneppen (2020) \cite{kirkegaard_optimal_2020}.} 
\label{fig:Kirkegaard1}
\end{center}
\end{figure}

Let us consider the vasculature pervading the islets of Langerhans of the pancreas, an assembly of cells producing and releasing hormones like insulin into the blood stream around them \cite{Berclaz.2016}. In a coarse-grained representation, the vasculature consists of a single inlet and a single outlet. Considering a disk-shaped network with the inlet and the outlet positioned at opposite sides, see Fig.~\ref{fig:Kirkegaard1}(a), minimizing for dissipation at a finite metabolic cost results in a single ``shunt'', see Fig~\ref{fig:Kirkegaard1}(b) along the shortest path between the inlet and the outlet. 
In the shunt architecture, hormones produced far-off the shunt do not enter the blood flow. To capture that the cells' product needs to be transported by the blood flow, a measure of transport efficiency might be the time it takes to travel along the flow from any cell in the network to the outlet \cite{kirkegaard_optimal_2020}. Identifying each network node $i$ as a producing cell and associated product travel time $T_i$ to the outlet, the function to optimize is the average travel time $\braket{T_i}$ across all nodes
\begin{align}\label{eq:AvgTimeNodeSink}
    \braket{T}=\frac{1}{N}\sum_iT_i,
\end{align}
where $N$ is the total number of nodes in the network. As the product is transported with the flow rate $Q_{ij}$ downstream toward the outlet, each node's travel time can be solved recursively by summing over the travel time of their next downstream nodes $j\in\mathcal{O}_i$ and the travel time along the tube connecting them $T_{ij}$ weighted by the flow rates along competing downstream routes,
\begin{align}\label{eq:T_i}
    T_i=\frac{\sum_{j\in\mathcal{O}_i}|Q_{ij}|(T_j+T_{ij})}{\sum_{j\in\mathcal{O}_i}|Q_{ij}|}
\end{align}
where $j\in\mathcal{O}_i$ is the set of nodes that are downstream from node $i$. Naively, one would evaluate the travel time  $T_{ij}$ along a tube $ij$ by the ratio of the length of the tube $l$ and the pervading cross-sectionally averaged flow velocity $\bar{u}_{ij}$, which is transformed into a condition on tube conductance $C_{ij}$ and flow rate $Q_{ij}$ as follows:
\begin{align}
    T_{ij}=\frac{l}{\bar{u}_{ij}}=\frac{l\pi a_{ij}^2}{Q_{ij}}=\sqrt{8\pi \mu l^3}\frac{\sqrt{C_{ij}}}{Q_{ij}},
\end{align}
 assuming Poiseuille flow. Given a fixed inflow at the inlet which equals the outflow at the outlet, and a fixed and constant tube length $l$, the travel time is trivially minimized by shrinking all tube radii to the lowest possible values, leading to very low conductances throughout the network. Diminishing tube conductance, however, comes at the cost of huge energy dissipation, scaling inversely with tube conductance, see Eq.~\eqref{eq:Dissipation}. Therefore, naive minimization of the travel time may not give physically meaningful predictions.
  Now, minimizing energy dissipation at a finite metabolic cost predicts how the optimal tube conductance scales with flow rate, see Eq.~\eqref{eq:OptConductance},
  \begin{align}
    C_{ij}\propto Q_{ij}^{\frac{2}{\gamma+1}}.
\end{align}
Therefore, using the optimal scaling of the tube's conductances in evaluating the travel time $T_{ij}$, travel time from all network nodes to the outlet and minimimal dissipation at a finite metabolic cost can be balanced \cite{kirkegaard_optimal_2020},
\begin{align}\label{eq:T_ij}
    T_{ij}\sim\frac{l}{|Q_{ij}|^{\frac{\gamma}{\gamma+1}}}.
\end{align}
Together Eqs.~\eqref{eq:AvgTimeNodeSink}, \eqref{eq:T_i}, and \eqref{eq:T_ij} define the optimization problem for transport in distributed production networks. Numerically minimizing the average travel time again yields, in general, multiple minima of a rugged cost function landscape. For $\gamma=1$, optimization leads to a periphery versus center network archetype, see Fig.~\ref{fig:Kirkegaard1}(c), with high conductance tubes spanning the periphery of the network branching out from the inlet and a high conductance tube along the center of the network extending to the outlet. This periphery - center architecture does indeed resemble the vasculature of pancreatic islets.

\begin{figure}[t]
\begin{center}
\includegraphics[width=\textwidth]{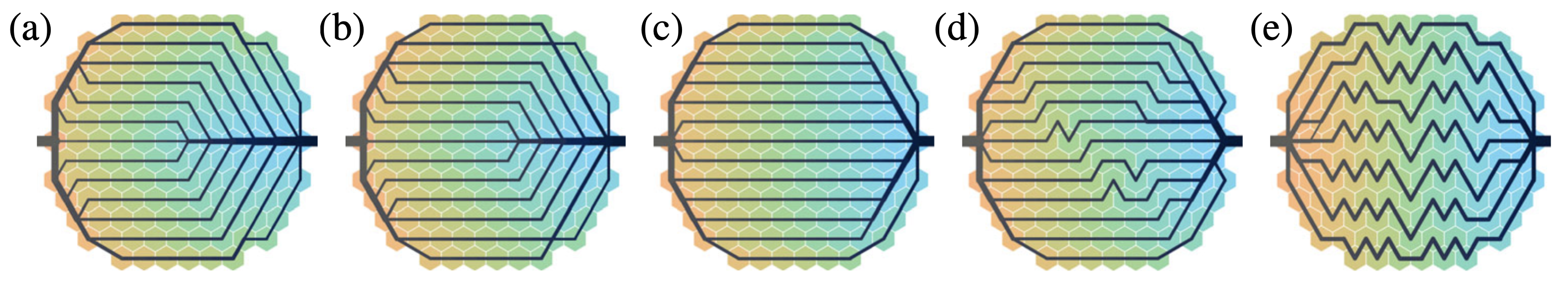}
\caption{\textbf{Optimal networks under minimal flow constraints.} Increasing minimal flow rate at any network node from (a) to (e) at (a) $Q_{\text{min}}= 0.042$ the network in Fig.~\ref{fig:Kirkegaard1}(c) the tube conductance of the periphery decreases. Then the extended outlet in (b) decreases as $Q_{\text{min}}$ is increased to 0.062 and to  $Q_{\text{min}}=0.073$ in (c). In (d) one horizontal connection has been removed by introducing kinks at $Q_{\text{min}}=0.078$, which increases the flow in the remaining connections. At yet increasing $Q_{\text{min}}=0.125$ (e) more horizontal connections are removed at the expense of longer, kinky connections. As $Q_{\text{min}}\rightarrow1$, the global optimum becomes a Hamiltonian path through the nodes. Reproduced from Kirkegaard and Sneppen (2020) \cite{kirkegaard_optimal_2020}.} 
\label{fig:Kirkegaard2}
\end{center}
\end{figure}
When considering optimal transport, another constraint to consider is the magnitude of the flow rate. Experimental studies of vasculature observe a uniform or at least minimal flow rate \cite{Chang.20175gc,Qi.2021,Qi.2024ryw}. For now, flow rates differ accross tubes in the periphery center architecture, see Fig.~\ref{fig:Kirkegaard1}. Despite the flow reaching all parts of the network, flow rates are, for example, larger in the center tube than in the tubes servicing it. Requiring a minimum flow rate $Q_{\text{min}}$ would constrain the flow through all nodes to be at least bigger than $Q_{\text{min}}$ \cite{kirkegaard_optimal_2020}
\begin{align}
    Q_i=\sum_{j\in\mathcal{O}_i}|Q_{ij}|\geq Q_{\text{min}}, \qquad \forall i
\end{align}
where $i$ denotes all the nodes in the network. Increasing the minimal flow rate $Q_{\text{min}}$ indeed has a drastic impact on the network architecture, first reducing the network radii hierarchy in the periphery, then in the center until only parallel connections between inlet and outlet persist, Fig.~\ref{fig:Kirkegaard2}. Increasing the minimal flow rate even further then reduces the number of parallel connections, thereby increasing the flow rate in each connection until only one long tube meanders through all network nodes. Note that for the single meandering tube, not only do all tubes have exactly the same flow rate, they also have exactly the same conductance. A result that is also obtained when optimizing minimal dissipation networks at a finite metabolic cost under a constraint of minimal variations in flow rate \cite{Chang.2019afm}. 

The example of distributed production networks nicely illustrates that the criterion of minimal dissipation at a finite metabolic cost is too narrow when it comes to the diversity of flow network architectures in biology. Yet, the ansatz of minimizing the average travel time suffers from one big conceptual draw back: it is a global cost function. A local read-out of information at individual tubes to optimize the global average travel time is unclear. Indeed, a minimal flow rate could be encoded locally, yet the travel time constraint is here the key to route the network through all network nodes. A constraint that fails under minimal flow rate only \cite{Chang.2019afm}. An alternative route is to consider the actual transport dynamics directly. 
\subsection{Transport and absorption in a single tube}
The transport of resources by fluid flow is governed by the advective transport along the streamlines of the fluid flow and the diffusion of the molecules independent of flow. Describing the resources by their concentration $c$, i.e.~the number of molecules per unit volume, the concentration evolves according to the advection-diffusion equation, here in cylindrical coordinates and laminar flow along a cylindrical tube,
\begin{align}\label{eq:Dispersion3D}
    \frac{\partial c}{\partial t}+u_z(r) \frac{\partial c}{\partial z}=\kappa\left[\frac{1}{r} \frac{\partial}{\partial r}\left(r \frac{\partial c}{\partial r}\right)+\frac{\partial^2 c}{\partial z^2}\right],
\end{align}
where $\kappa$ denotes the molecular diffusivity of the resource molecules. For Poiseuille flow in a tube of radius $a$ and cross-sectionally average flow velocity $\bar{u}$ the flow velocity driving advection is given by $u(r, z)=2\left(1-r^2 / a^2\right)\bar{u}$.
 The absorption of resources at the tube wall is represented by the following boundary condition:
\begin{align}
    \left.\kappa \frac{\partial c}{\partial r}\right|_{r=a}+\nu c(a,z,t)=0,
\end{align}
where $\nu$ denotes the metabolite absorption rate at the tube wall. While there is no closed analytical solution for $c(r,z,t)$ in  Eq.~\eqref{eq:Dispersion3D}, the advection-diffusion equation can be approximated by the dynamics of the cross-sectionally averaged concentration $\bar{c}(z,t)$ \cite{Taylor.1953,Aris.1956}. This Taylor approximation is valid if the time to average out the cross-sectional variations in molecule concentration by diffusion is shorter than the time for advection along the tube, i.e.~$a^2/\kappa\ll l/\bar{u}$. Also, absorption at the tube walls further introduces gradients in molecule concentration which are averaged out by diffusion if the time for diffusion is quicker than the time scale over which absorption generates gradients, i.e.~$a^2/\kappa\ll a/\nu$. In these limits, one can apply the heuristic Taylor dispersion derivation to the case of absorption at the tube wall to arrive at \cite{meigel_flow_2018},
\begin{align}\label{eq:Dispersion1D}
    \frac{\partial\bar{c}}{\partial t}= -\frac{2 \kappa}{a^2} \frac{4 \nu a/\kappa}{4+\nu a/\kappa}\bar{c}-\frac{12+\nu a/\kappa}{12+3 \nu a/\kappa}\bar{u} \frac{\partial\bar{c}}{\partial z}  +\left(\kappa+\frac{12+\nu a/\kappa}{12+3 \nu a/\kappa} \frac{\bar{u}^2 a^2}{48 \kappa}\right) \frac{\partial^2\bar{c}}{\partial z^2}.
\end{align}
We readily read off, that the concentration of molecules decays over time, gets advected, and diffuses. The effective advection velocity is larger than the cross-sectionally averaged flow velocity $\bar{u}$ since absorption removes slow molecules close to the tube wall, thereby effectively increasing the cross-sectionally averaged velocity of transport. Also, the diffusivity is enhanced as the radial gradient in flow velocity between streamlines enhances the longitudinal spread of molecules that are diffusively hopping between streamlines. Note, better approximations to any desired order can be derived using the center manifold approach \cite{Marbach.2019} that can even be compared to exact solutions of the decay of concentration under absorption, advection and diffusion \cite{Lungu.1982}. The strength of the heuristic approximation above is that it allows for physical intuition into the competing factors driving the absorption of molecules. The steady state of the cross-sectionally averaged concentration along a tube of length $l$ is an exponential decay from upstream concentration $\bar{c}_0$\cite{meigel_flow_2018}:
\begin{align}
    \bar{c}(z)=\bar{c}_0 \exp \left(-\alpha \frac{z}{\ell}\right), \qquad \text{where }\alpha=\frac{24 \cdot \mathrm{Pe}}{48+\nu a/\kappa \cdot \mathrm{Pe} / \mathrm{S}}\left(\sqrt{1+8 \cdot \frac{S}{\mathrm{Pe}}+\frac{3}{4} \cdot \nu a/\kappa}-1\right)
\end{align}
where the characteristic decay rate is goverened by three non-dimensional parameters, $\nu a/\kappa$, the ratio of diffusion to absorption time scale, the Péclet number $\mathrm{Pe}=\bar{u}l/\kappa$, which is the ratio between diffusive and advective time scales, and the Damköhler number $\mathrm{S}=\nu l / a\bar{u}$, which is the ratio between the timescale for absorption $\nu/a$ and the time to be advected out of the tube $l/\bar{u}$. Recall, that $\nu a/\kappa\ll 1$ to fullfil the Taylor dispersion approximation, rendering the Péclet $\mathrm{Pe}$ and the Damköhler number $\mathrm{S}$ to be the only freely varying non-dimensional parameters.

The heuristic derivation also accounts for a direct result of the concentration $c(r,z,t)$ and thus permits the derivation of the overall flux of molecules absorbed along a tube $\phi$, by integrating over the tube wall $\mathcal{W}$ extending along the length $l$ of the tube, $\phi=2 \pi a \int_{\mathcal{W}} \kappa \frac{\partial c}{\partial r} \mathrm{~d} z$. The flux of absorbed molecules is to first order a function of the amount of influx of molecules $J_0=\bar{c}_0(\bar{u}+\kappa \alpha/l)$ being advected or diffusing into the tube. Therefore, specifying the absorption capacity $\hat{\phi}$ as the ratio of total absorption flux relative to concentration influx is meaningful: $\hat{\phi}=\phi/\pi a^2 J_0$ with $0\leq\hat{\phi}\leq1$. The absorption capacity spans three different regimes as a function of the Péclet number $\mathrm{Pe}$ and the Damköhler number $\mathrm{S}$ \cite{Meigel.2019}. At $\mathrm{S}\gg 1$ and $\mathrm{S}\gg1/\mathrm{Pe}$, the decay exponent $\alpha$ is large and every molecule entering the tube gets absorbed. In contrast, $\alpha$ is small in both the diffusive regime at $\mathrm{Pe}\ll \mathrm{S}$ and $\mathrm{Pe}\ll1/\mathrm{S}$, and the advective regime at $\mathrm{S}\ll 1$ and $\mathrm{S}\ll\mathrm{Pe}$, where the flux of absorbed molecules is roughly given by $\phi=2\pi a l \nu \bar{c}_0$, although due to different physical mechanisms at play \cite{Meigel.2019}. In the most important regime for physiological application, the advective regime, where molecules are mainly advected by flow along the tubes rather than diffusion, the absorption capacity $\hat{\phi}$ takes a simple expression of the form \cite{meigel_flow_2018}
\begin{align}\label{eq:AbsconductanceExpression}
    \hat{\phi}=\frac{2 \nu l}{a\bar{u}+2 \nu l}.
\end{align}
This closed analytical expression gives the immediate insight that large flows impede absorption as the molecules travel through the tube way too fast to get absorbed. Notably, the framework of molecule transport and absorption in a single tube is the building block to address the network architecture that optimizes supply. 
\begin{figure}[t]
\begin{center}
\includegraphics[width=\textwidth]{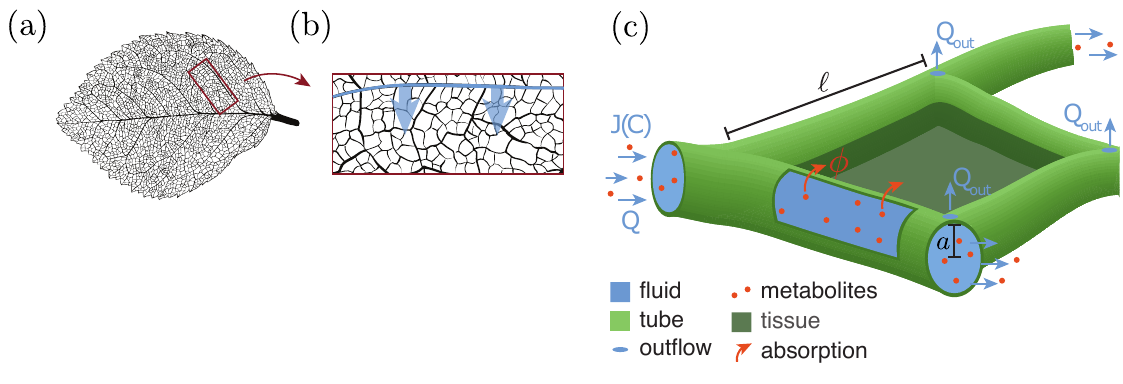}
\caption{\textbf{Schematic sketch of resource supply in leaves.} (a) Vasculature of a leaf displaying the primary vein horizontally at the center and secondary veins as next biggest veins departing from the primary vein, down to the highly inter-webbed higher order veins. (b) The secondary vein (thick blue) supplies the tubular higher order vein network with resources and fluid. (c) Xylem vessel network modeled as network of tubes of varying radius $a$ and fixed length $l$. Inflow at fluid flow rate $Q$ and resource flux $J$ from upstream tubes (left). Fluid evaporation through stomata at the leaf surface modeled by constant outflow $Q_{\mathrm{out}}$ at every network node. Resource molecules are advected and diffuse within the fluid. In addition, molecules get absorbed $\phi$ along the tube wall into tissue at a constant rate $\nu$. Reproduced from Meigel and Alim (2018) \cite{meigel_flow_2018}. } 
\label{fig:Meigel1}
\end{center}
\end{figure}
\subsection{Optimizing networks for homogeneous supply}

We revisit the venation of leaves as an example. However, now we zoom into the finely inter-webbed higher order venation in between the bigger so-called secondary veins, see Fig.~\ref{fig:Meigel1}. 
Here, the fluid enters from the secondary vein into the higher order venation, where the fluid evaporates through stomata. Therefore, there are multiple inlets along the secondary veins servicing outlet represented by the nodes throughout the higher order venation. As resources that are essential for cell maintenance are transported within the stream of flow , the question is, which flow and network architecture characteristics are required to ensure homogeneous supply of resource molecules transported by the flow.

\begin{figure}
\begin{center}
\includegraphics[width=0.9\textwidth]{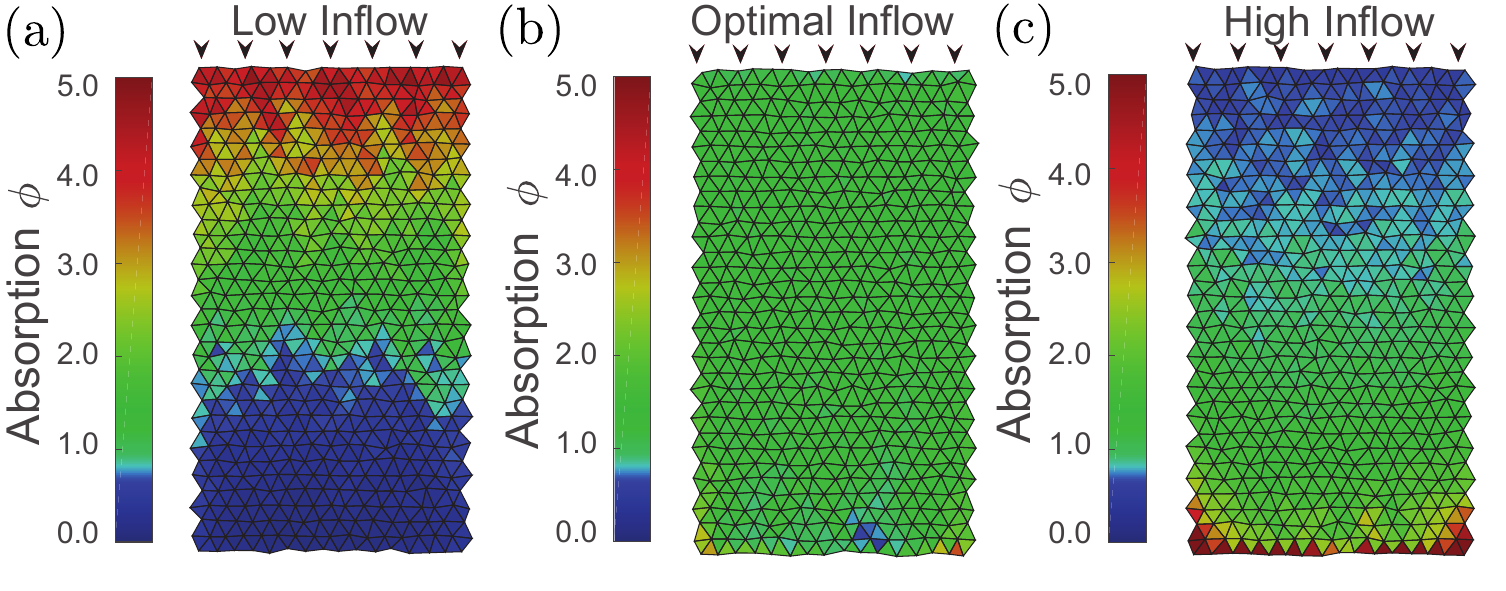}
\caption{\textbf{Supply patterns are controlled by fluid inflow rates.} Supply pattern of a rectangular tissue section pervaded by a transport network for increasing fluid inflow rate ranging from (a) $Q_{in}=0.8\times 10^{-6} \text{mm}^3\text{s}^{-1}$, via (b) $Q_{in}=3.2\times 10^{-6} \text{mm}^3\text{s}^{-1}$,to(c) $Q_{in}=6.4\times 10^{-6} \text{mm}^3\text{s}^{-1}$. Resource molecules are absorbed across tube walls into the tissue. (a,b,c) Supply pattern in every triangulated tissue section given by the normalized average steady state flux of absorbed resource molecules along neighboring tubes. At low inflow rates (a) resources are absorbed close to inflow and are not transported through the network while for high inflow rates (c) resources get flushed through the network for being absorbed mainly at the end.  In between these two cases, an optimal inflow rate with the lowest variance in flux of absorbed molecules exists (b). Reproduced from Meigel and Alim (2018) \cite{meigel_flow_2018}. } 
\label{fig:Meigel2}
\end{center}
\end{figure}

Describing the flow along the higher order venation network as Poiseuille flow, the framework of Eqs.~\eqref{eq:Dispersion1D} and \eqref{eq:AbsconductanceExpression} describes the supply of resources as the flux of absorbed resource molecules through the tube walls. Numerically solving for the steady state of supply across a network of tubes of uniform radii, the impact of the inflow rate becomes apparent, see Fig.~\ref{fig:Meigel2}. At low inflow, the tubes close to the inlets receive a lot of supply and exhaust the molecules such that those tubes further off receive nothing anymore. At high inflow, on the contrary, the molecules are flushed past the tubes close to the inlet so fast that almost no absorption is taking place, and only tubes further on, where flow velocity is reduced due to the evaporation of fluid along the way, receive a lot of supply. In between these two extremes, there is a flow rate which provides homogeneous supply across all tubes in the network. 

\begin{figure}[ht]
    \centering
    \begin{tikzpicture}
        \draw (0,0) -- (2,0);
        \draw (0,-0.4) -- (2,-0.4);
        
        \draw (3,0) -- (5,0);
        \draw (3,-0.4) -- (5,-0.4);
        
        \draw (6,0) -- (8,0);
        \draw (6,-0.4) -- (8,-0.4);
        
        \fill (8.6,-0.2) circle (0.05);
        \fill (9,-0.2) circle (0.05);
        \fill (9.4,-0.2) circle (0.05);
        
        \draw (10,0) -- (12,0);
        \draw (10,-0.4) -- (12,-0.4);
        
        \node[below] at (1,-0.4) {1};
        \node[below] at (4,-0.4) {2};
        \node[below] at (7,-0.4) {3};
        \node[below] at (11,-0.4) {M};
        
        \draw[->] (-0.8,-0.2) -- (0,-0.2) node[midway, above] {$Q_{\text{in}}$};
        
        \draw[->] (2.5,-0.2) -- (2.5,0.5) node[right] {$Q_{\text{out}}$};
        \draw[->] (5.5,-0.2) -- (5.5,0.5) node[right] {$Q_{\text{out}}$};
        \draw[->] (12.5,-0.2) -- (12.5,0.5) node[right] {$Q_{\text{out}}$};
    \end{tikzpicture}
    \caption{    \label{fig:1Dtransport}\textbf{Schematic representation for the derivation of optimal inflow rate} One-dimensional network of $M$ connected tubes at inflow $Q_{\text{in}}$  in the first tube and subsequent outflow $Q_{\text{out}}$ at each node. The resource molecules are transported along with the flow, except that they cannot escape at the nodes. They are either absorbed in the tube walls or are transported to the next tube.}
\end{figure}
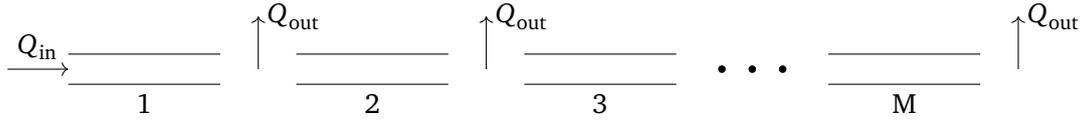
To derive the optimal flow rate that yields uniform supply, we reduce the interlaced network to a one-dimensional series of $M$ identical tubes of radius $a$ and length $l$, with their flow rates $Q_m$ for the $m$-th tube, see Fig.~\ref{fig:1Dtransport}. Resource molecules are transported by the flow which is flowing into the first tube at a rate $Q_{\text{in}}$. At each node between two tubes, there is a constant outflow $Q_{\text{out}}$, which matches in sum the inflow, i.e.~$Q_{\text{out}}=Q_{\text{in}}/M$. Due to the evaporation modeled as outflow at each network node, the flow rate decreases the further away the tube is from the inlet. Therefore, the flow rate of subsequent tubes and their cross-sectional average flow velocity is:
\begin{align}
    Q_{m+1}&=Q_m-Q_{\text{out}},\nonumber \\
    \implies \bar{u}_{m+1}&=\bar{u}_m-\frac{Q_{\text{out}}}{\pi a^2}=\bar{u}_m-\frac{Q_{\text{in}}}{M\pi a^2}.
    \label{eq:subsequentFlowLeaf}
\end{align}
Resource molecules do not escape at network nodes, they do not evaporate, they either get absorbed or are transported to the next tube. As a result, from the total amount of molecules entering a tube $m$, a fraction $\hat{\phi}$ is absorbed and the remaining fraction $(1-\hat{\phi})$ enters tube $(m+1)$. Thus, the flux of  absorbed molecules  in tube $m$ is given by
\begin{align}
    \phi_m = \pi a^2 J_0 \hat{\phi}_m\prod_{j=1}^{m-1}(1-\hat{\phi}_j).
\end{align}
At homogeneous absorption, $\phi_m$ is the same $\forall m$, which means that also the flux of absorbed molecules at subsequent tubes is the same, i.e., 
\begin{align}\label{eq:FluxConsec}
    \prod_{j=1}^{m-1}(1-\hat{\phi}_j)\hat{\phi}_m&=\prod_{j=1}^{m}(1-\hat{\phi}_j)\hat{\phi}_{m+1} \nonumber \\
    \implies\hat{\phi}_m&=(1-\hat{\phi}_m)\hat{\phi}_{m+1}
\end{align}
Substituting the value of absorption capacity $\hat{\phi}_m$ from Eq.~\eqref{eq:AbsconductanceExpression}, we get
\begin{align*}
\frac{2\nu l}{a\bar{u}_m+2\nu l}
&= \frac{a\bar{u}_m}{a\bar{u}_m+2\nu l}\cdot \frac{2\nu l}{a\bar{u}_{m+1}+2\nu l} \qquad \left(\because 1-\hat{\phi}_m = \frac{a\bar{u}_m}{a\bar{u}_m+2\nu l}\right)\\
\implies 1 &= \frac{a\bar{u}_m}{a\bar{u}_{m+1}+2\nu l},\\
\implies \bar{u}_{m+1} &= \bar{u}_m - \frac{2\nu l}{a}.
\end{align*}
Comparing with Eq.~\eqref{eq:subsequentFlowLeaf} and equating the decrements, we find that there is an optimal inflow rate $Q_{\text{in}}^*$ which permits homogeneous absorption:
\begin{align}
    Q_{\text{in}}^*=2\pi a lM \nu,
\end{align}
which is in quantitative agreement with the numerical simulations \cite{meigel_flow_2018}. The inflow rate matches the integrated absorption flux approximated by the wall surface of all tubes $2\pi a lM$ times the absorption rate $\nu$. Numerically adjusting tube radii at non-optimal inflow rate compensates by reducing tube radii and thus tube wall surface where supply is too high and conversely increases wall surface at tube walls where supply is below average \cite{meigel_flow_2018}. The biggest lever next to the inflow rate is therefore the tube radius. Therefore, local read-out of resource availability and in response local adaptation in tube radius can adjust for local resource demand \cite{Meigel.2019} and homogeneous supply \cite{Bouvard.2024}. Note, however, that homogeneous supply into a tissue by adjacent tubes can also be locally regulated by the tissue instead of the tube radii \cite{Gavrilchenko.2021}. Here, tissues inflate proportionally to the amount of resources supplied by the tubes and balance their size across a tissue \cite{Gavrilchenko.2021}.

Optimization for homogeneous supply is in stark contrast to optimization for energy dissipation at a finite metabolic cost. At least when considering the typically physiological metabolic scaling of $\gamma=1/2$ in Eq.~\eqref{eq:GamLam}, where tree-like network architecture minimizes energy dissipation. In contrast, optimization for homogeneous supply favors a finely reticulated networks \cite{Kramer.2023}. When optimizing both at the same time, the balance of weights in the cost function determines the dominant network architecture \cite{Kramer.2023}.
\section{Conclusion}
Without fluid flow, life as we know it would not exist. Fluid flow is a key player in the function of biological systems. Yet, understanding the dynamics of flows is challenging as fluid flows are inherently coupled over many spatial and temporal scales. The fluids themselves transport - possibly active - molecules, are visco-elastic, couple strongly to their surrounding walls, or take place in geometrically complex spaces. Placed into biological systems, the dynamics of fluid flow can be very complex, as flows impact morphology, biological signaling, and the dynamics of development. Here, we focused on flow networks in biology to give a concrete idea of how the physics of fluid flows and the self-organization of biology merge into complex architecture and function, and yet, can be dissected into conceptual ideas that only require a single equation. Understanding flow networks in biology is an amazing example where both top-down principles and bottom-up physical equations of motion meet to generate insights into the physical principles that make life work. We hope that this example not only demonstrates the reason for studying biological physics, but also inspires the reader to devote their own time to pondering the role of flows in biology.
\section*{Acknowledgements}
Thanks goes to the organizers of the 2023 Les Houches Summer School on Theoretical Biological Physics and my collaborators on my journey through flow networks in biology. 


\paragraph{Funding information}
This work was supported by DFG through AL 1429/6-1 as well as funding by the IGSSE / TUM Graduate School for IPT SOFT. This work received funding from the European Research Council (ERC) under the European Union’s Horizon 2020 research and innovation program (grant agreement No. 947630, FlowMem). 



\end{document}